# Enerzyme: A Framework for Efficient Training of Reactive Neural Network Potentials for Enzyme Catalysis with Application to Methyltransferases

Weiliang Luo[1,2] and Heather J. Kulik[1,2,*]

[1]*Department of Chemical Engineering, Massachusetts Institute of Technology, Cambridge, MA 02139, USA*
[2]*Department of Chemistry, Massachusetts Institute of Technology, Cambridge, MA 02139, USA*
*corresponding author email: hjkulik@mit.edu

ABSTRACT: Quantum mechanical (QM) cluster models provide an effective framework for mechanistic studies of enzymatic reactions but remain computationally demanding. Neural network potentials (NNPs) offer a promising route to reduce their cost, but enzymes introduce new challenges not faced by NNPs for small molecules, including large system sizes, implicit-solvent environments, substantial polarization, and charge transfer. Here, we demonstrate an integrated software framework for efficient training of NNPs for mechanistic study of enzymes with demonstrations on QM cluster models of *S*-adenosyl-L-methionine-dependent methyltransferases (MTases). Our Enerzyme code introduces modular electrostatics-aware NNP architectures and combines automated QM-cluster construction with reactive dataset generation. The Enerzymette subpackage also automates reaction pathway exploration at both NNP and DFT levels of theory. We show that iterative flexible scans and nudged elastic band calculations impose stricter requirements on NNPs than conventionally employed dataset metrics. Nevertheless, we develop an approach to train NNPs on fewer than 1,000 system-specific datapoints to reproduce reaction energetics and transition state structures on MTase clusters containing up to 545 atoms to approaching chemical accuracy of the reference. We show that direct supervision of atomic charges and alignment of dielectric screening with QM-cluster calculations substantially improve simulation stability and accuracy and that multitask-learned atomic charges predict correct charge transfer and polarization trends and provide chemically meaningful descriptors of reactivity. Furthermore, we demonstrate transferability across chemically diverse catechol *O*-methyltransferase substrates, indicating NNPs learn generalizable reactivity patterns as training data is expanded across multiple enzymes. Together, these results establish a foundation for accelerating enzyme mechanistic studies and provide insights for future NNP development for biomolecular reactivity.

## 1. Introduction.

Enzymes catalyze biochemical transformations efficiently and selectively[1,2] through interactions between the reacting species and the surrounding protein environment[3-5]. As a result, simulations that aim at understanding enzymatic catalysis must capture both the reacting species and the influence of the protein environment[6,7]. Due to computational constraints in large biomolecules, practical approaches typically include a multi-scale quantum mechanical (QM) description of the core active site combined with treating the remaining system at a lower level of theory such as a molecular mechanics (MM) force field. QM-cluster models represent a widely used and complementary alternative to QM/MM embedding that omits the atomistic details of the MM environment and avoids the need for extensive sampling[8-10]. Besides providing valuable mechanistic understanding in case studies[11-13] thanks to increasingly large QM cluster models comprised of hundreds of atoms, recent advances in QM cluster workflow automation have further facilitated systematic comparison across different mutants[14] and enzyme families[15,16].

Importantly, previous studies have shown that converged descriptions of enzymatic reactivity often require sufficiently large QM regions in embedding schemes[17,18]. Since the use of minimal theozymes consisting of reactants and catalytic groups[19], modern QM cluster models have expanded to 200–600 atoms to include other cofactors and surrounding residues for their long-range interactions and second-shell effects[20,21]. Nevertheless, even with GPU-acceleration of quantum chemistry[22,23] at the protein scale[24-26], extensive explorations of complex potential-energy surfaces (PESs) in large QM-cluster models remain computationally demanding, limiting the throughput of simulations for large-scale, comprehensive mechanistic studies. Neural network potentials (NNPs)[27-29] have emerged as a surrogate modeling tool of chemical reactions[30-32] with near-quantum-chemical accuracy at dramatically reduced computational cost[33-35]. In recent years, rapid advances in NNP foundation models[36] have extended their application in catalysis research from small molecules[37-39] to enzymes[40-42]. These developments suggest a compelling opportunity to accelerate mechanistic studies of enzyme catalysis[41,43]. Progress in this direction will also inform the development of ML/MM approaches[42,44,45] or NNP-driven simulation of the full system for small-to-medium sized reactive biomolecules[46-48].

However, enzyme QM clusters present new challenges to most NNP formulations for molecular systems[49-51]. Enzyme clusters are significantly larger than small molecules in typical reaction datasets[52-54], and thus it is harder to learn their conformational and configurational space from training data[55]. QM cluster calculations are often embedded in implicit solvents[5,56] to represent the dielectric of the surrounding protein environment, blocking the direct transfer of the vacuum PES learned by foundation models[57,58]. In addition, many enzymatic reactions involve substantial polarization and charge transfer[18,59,60], and thus, electrostatics must be well-described throughout the reaction pathway[61-63]. Beyond the PES, electronic descriptors such as atomic charges are also frequently used to interpret interaction patterns and reactivity trends[64-66], which are present in only a few NNP architectures[67-69]. NNP-driven structure–activity explorations, also require robust evaluation of flexible scans and nudged elastic band (NEB) workflows[70-72], which have not been extensively tested in enzymes. Mechanistic studies for enzyme cluster models introduce new demands on both the electrostatic-aware development[73-75] and workflow-level evaluation of NNPs[76,77] that link prediction and simulation with reactivity trends and mechanism.

As a model system to demonstrate potential challenges in NNP development for enzymatic reactions, we focus on the *S*-adenosyl-L-methionine (SAM)-dependent methyltransferases (MTases) enzyme family that catalyze the $S_N2$ transfer of a methyl group to a wide array of nucleophiles. MTases constitute one of the most widespread enzyme superfamilies in nature[78,79] and participate in gene regulation[80], protein repair[81], drug metabolism[82], and natural product biosynthesis[83]. Our previous QM studies have shown that electrostatic interactions and charge transfer play a central role in their catalytic behavior[17,18,64], making them particularly informative test systems for electrostatics-aware NNPs. Furthermore, the broadly conserved reaction pattern of MTases[84] enables systematic comparisons across chemically diverse enzyme systems.

To both identify and address gaps in NNP development for enzyme catalysis, we present an integrated framework for NNP-driven mechanistic simulations demonstrated on QM cluster models of MTase enzymes. We combine automated QM-cluster construction and reactive dataset generation with electrostatics-aware NNP architecture and workflow-level structure–reactivity exploration. We investigate whether NNPs can reproduce key energetic, structural, and electronic features of QM-level mechanistic studies, further examine whether NNP-derived electronic descriptors capture reactivity trends, and assess the emergence of transferable biochemical modeling as training data expand to more enzyme systems. Together, these results establish a foundation for accelerating mechanistic studies of enzymatic reactions while preserving key chemical insights from quantum chemistry.

## 2. Computational Details.

### QM cluster model preparation.

The QM cluster models of enzymes PfPMT[85] and HcgC[86,87] were obtained from our previous QM/MM studies[64] on methyltransferases (Supporting Information Table S1 and S2). QM cluster models of COMTs from the Protein Data Bank (PDB IDs: 1VID[88], 2CL5[89], 2ZVJ[90], 3S68[91], 4XUD[92], 4XUE[92], 5LSA[93], 6LFE[94], 7XJB[95]) surrounding their substrate, SAM cofactor, and $Mg^{2+}$ ion were built with the QuantumPDB package[16] (Supporting Information Table S3 and S4).

**Reactive sampling of cluster models.**

We conducted steered MD in the atomic simulation environment (ASE)[96] with its native xTB[97] and PLUMED[98,99] wrapper. Similar to ref. 37, a moving bias potential was applied to the collective variable (CV) $d_{CX}$–$d_{SC}$, steering the reactant to the product region. One-sided harmonic restraint potentials were added to all pair-wise distances between protein backbone atoms, water oxygens, and ions (See Supporting Information Table S5 for parameter details). The underlying level of theory was GFN1-xTB[100] with analytical linearized Poisson–Boltzmann (ALPB) implicit solvation[101], while UMA-s-1.1 NNP[102] in the fairchem package was also used for comparison with GFN1-xTB (See Supporting Information Figure S2).

**DFT calculations in QM cluster models.**

All *ab initio* DFT energies and forces used in this simulation were calculated by the GPU-accelerated quantum chemistry code, TeraChem v1.9[22,103-105], at the B3LYP/6-31G*[106,107] level of theory with DFT-D3 dispersion correction[108] using default Becke–Johnson (BJ) damping[109]. A conductor-like screening model (C-PCM) for implicit solvation[110,111] with a dielectric constant of 10 was applied to the QM cluster to approximate the protein's electrostatic environment. All 1.2×CM5[112,113], Hirshfeld[114], ADCH[115], RESP[116], and VDD[117] atomic charges used in this study were calculated based on the Molden files obtained from TeraChem output artifacts with Multiwfn[118,119]. We use the same charge-transfer analysis method for MTases as in our previous QM/MM studies[17,18,64].

**Structure–reactivity study in cluster models with DFT and NNP.**

Flexible scans of the clusters were run along the $d_{\mathrm{CX}}$ reaction coordinate, starting from an optimized reactant geometry to an anticipated $d_{\mathrm{CX}}$ at the product state incrementally by 25 steps (Supporting Information Table S6). For all geometry optimization involved in flexible scans, DFT calculations used the geomeTRIC optimizer[120] in TeraChem, and the wave function from the previous optimization step was used as the initial guess for the SCF to improve convergence to a smooth reaction coordinate, while NNP calculations used the L-BFGS optimizer[121] in ASE with our Enerzyme simulate submodule. The maximum force convergence criterion was set to $4.5\times10^{-4}$ hartree • $\text{Å}^{-1}$ for both DFT and NNP optimizations.

Nudged elastic band (NEB) calculations of the clusters were run between pre-defined reactant and product structures. Initial images between the optimized reactant and product endpoint structures were interpolated with the image-dependent pair potential (IDPP) method[122,123]. The path was then optimized with the climbing-image NEB (CI-NEB) method[124]. We interfaced ORCA 6.1.0[125-127] as an external NEB optimizer[128] with TeraChem as the DFT energy/forces supplier by our Enerzymette `orca_terachem_request` submodule or Enerzyme as the NNP energy/forces supplier. For NNP calculations, we built a server-client mode to reduce the overhead of module loading and neural network initialization. That is, our Enerzyme `listen` submodule deployed a server for energy and forces calculation with a trained NNP, and our Enerzyme `request` submodule invoked the calculation and parsed the results for the ORCA NEB optimizer.

During all geometry optimizations, flexible scans, and NEB calculations in the enzyme cluster model, protein backbone heavy atoms were fixed at their original positions to preserve the structural influence of the protein backbone. To minimize the effects of initial conditions on the optimization results, both the DFT and NNPs were initialized from the same DFT-optimized endpoints required for flexible scans and NEBs. For NNP calculations, iterative scans and NEBs were fully automated by our Enerzymette `enerzyme_scan` and `enerzyme_neb` submodules. For DFT calculations, due to the substantial wall time per iteration, these workflows were manually managed using the same criterion.

**NNP architecture and training.**

In this work, we used the message-passing GNN cores of PhysNet[129], SpookyNet[130], and MACE[131] NNPs with our customized pre-core and post-core layers (Supporting Information Tables S7–S9). They were implemented in our Enerzyme package with the PyTorch[132] deep learning framework. The parameters of NNPs were optimized by minimizing the loss function with the AMSGrad variant of the Adam optimizer.[133] During training, we used a linear learning rate scheduler with warmup and determined the termination of training by early stopping[134]. At each epoch before reaching the maximum epoch number, a judge score was evaluated on the validation set using NN parameters obtained from an exponential moving average (EMA)[135], and the training was terminated if it hadn't improved for a patience interval of epochs. The judge score shared the same weighted-sum form and weights on output properties as the loss function (equation 1):

$$m = w_E m_E + w_{\boldsymbol{F}} m_{\boldsymbol{F}} + w_\mu m_\mu + w_Q m_Q + w_q m_q, \quad (1)$$

where $m_x$ is a metric term for a property $x$ (Supporting Information Table S10). After training, the model with the best judge score on the validation set is selected for downstream use. The whole training procedure was managed by the Enerzyme `train` submodule.

## 3. Results.

### 3a. Standardized model enzyme construction and reactive NNP data generation.

We construct QM cluster models of MTase active sites and build their NNP datasets from efficient reactive event sampling with QM labels to study enzymatic reactions in MTases with DFT and NNP approaches (Figure 1). For this purpose, QM cluster construction needs to be reproducible and automated across different systems. For each QM cluster, we also need to sample the relevant PES in terms of both the reaction coordinate and other orthogonal degrees of freedom to train an accurate and stable NNP. Therefore, we first harnessed our code QuantumPDB[16] to build QM clusters of MTases from the Protein Data Bank (PDB) around their reaction centers, including the cofactor SAM, the $Mg^{2+}$ ion[136] in catechol *O*-methyltransferases (COMT), and a range of substrates, alongside QM clusters that we extracted from our previous QM/MM studies[64] (see Computational Details). Then, we tailored an enhanced sampling protocol for MTase clusters to cover the entire reaction pathway while simultaneously generating diverse configurations of the surrounding catalytic environment.

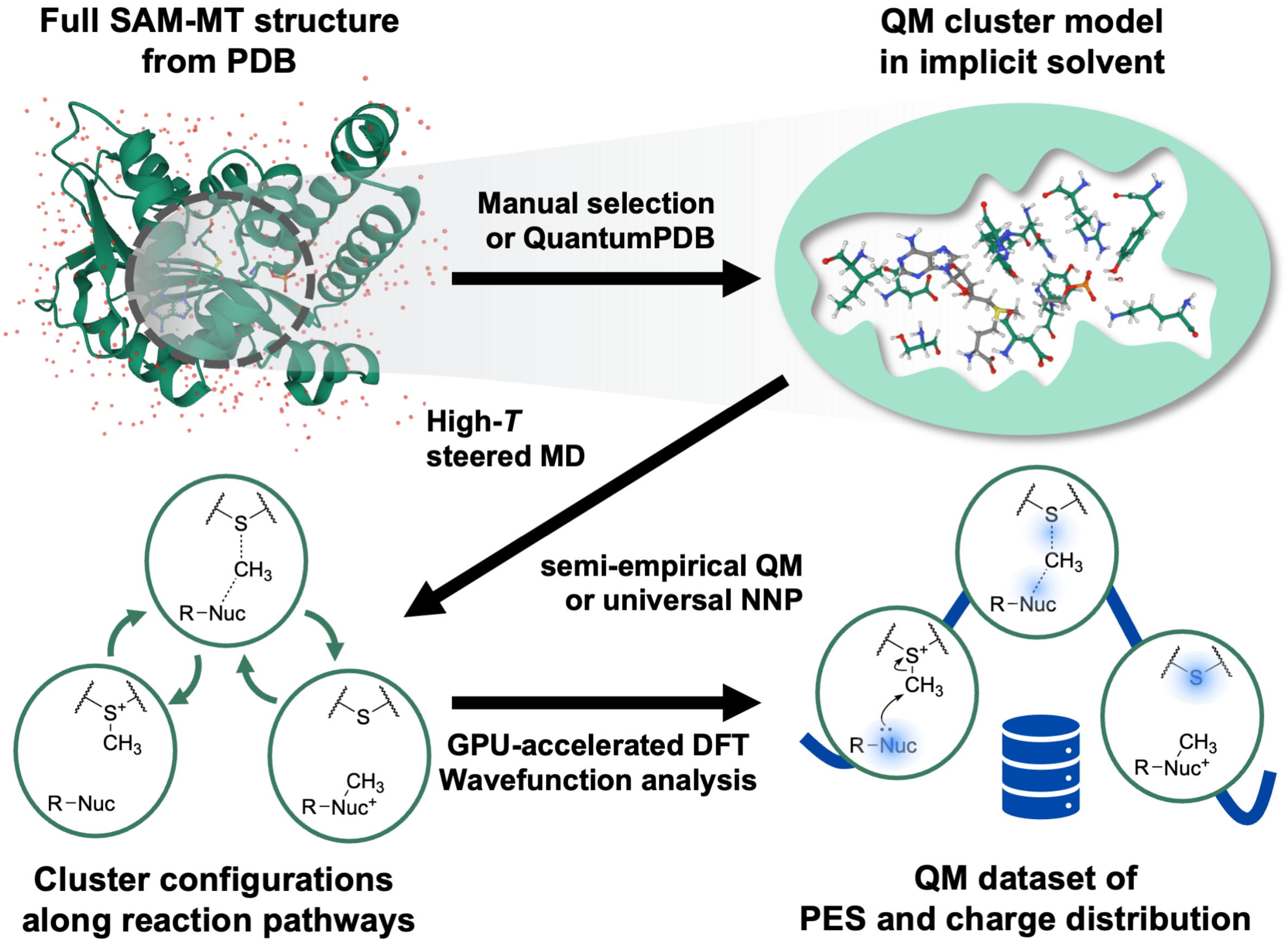

**Figure 1.** An overview of data generation for SAM-MT enzyme cluster models. From a full SAM-MT structure in the PDB, we construct its cluster model by manual selection or an automated workflow with QuantumPDB. We use high-temperature steered MD to sample the configurations along reaction paths with efficient, reactive simulators, including semi-empirical QM methods or universal NNPs. The obtained structures are labeled by DFT with the implicit solvent model for their potential energy surface and charge distribution information.

In our reactive sampling protocol, steered MD[137] is employed to overcome reaction barriers and enhance the rare reactive events. To sample orthogonal degrees of freedom, especially including molecular conformations and non-covalent interactions, we use a high temperature of 500–1000 K during the steered MD run to promote their exploration. To maintain a suitably compact model of the enzyme active site throughout the simulation while allowing sufficient conformational flexibility, soft restraints are applied between different molecular fragments (see Computational Details). This approach yields a stable trajectory with uniform data distribution across the reactant, transition, and product regions (Supporting Information Figure S1). These snapshoots from high-temperature MD can in turn be labeled with the same electronic structure method as in the reference QM cluster to obtain energies, forces, dipoles, and atomic charges in the NNP dataset (see Computational Details).

The underlying interatomic potential of the reactive sampling must be able to describe bond breaking and bond formation, remain stable under high-temperature steered MD, and efficiently compute energies and forces for large-scale simulations. As an initial choice, we employed the xTB family of semi-empirical quantum chemistry methods[100,138], which provides a good balance between reactivity description and computational efficiency. In practice, GFN1-xTB enables stable simulations at 500–1000 K across multiple SAM-MT clusters and, with CPU parallelization, can complete 20,000 MD steps for a ~300-atom cluster within 4 hours. However, xTB methods encounter difficulties when describing highly charged molecules, such as polyphosphate groups, leading to abnormal molecular geometries or instability in electronic structure calculations in some systems. During the course of this study, universal pretrained NNPs for reactive molecular systems became available. Indeed, we found that foundation models such as UMA[102] can more efficiently and accurately simulate reactive systems where xTB performs poorly, offering an alternative for robust and efficient sampling (Supporting Information Figure S2).

Overall, given structure of an enzyme-substrate complex, a specification on the substrate identity, a reactive interatomic potential, and an electronic structure code, our MTase model construction and NNP generation represent a standardized and theory-agnostic procedure for the whole MTase family.

**3b. NNP frameworks for mechanistic simulations in model enzymes.**

Performing QM-cluster-like simulations with NNPs in a manner that can supplant DFT introduces new challenges due to the large size, structural complexity, and diverse physical information contained in enzyme QM cluster models. On one hand, such an NNP must not only fit the potential energy surface (PES) to drive molecular simulations, but to represent a true replacement of a DFT calculation, it must also model long-range electrostatics, describe distributions of partial charges, and distinguish systems of different total charge. Furthermore, evaluating the NNP's reliability and stability for enzyme mechanistic studies in comparison to DFT calculations also requires testing on realistic QM-cluster simulation workflows, such as flexible scans and NEB calculations. To address these two challenges, we introduced two complementary frameworks: Enerzyme, an NNP development, training, and deployment platform that enables electrostatics-aware NNP simulations tailored for QM cluster models; and Enerzymette, a workflow manager that standardizes iterative structure–reactivity exploration on complex enzyme cluster PESs and aligns NNP evaluations to identical DFT reaction-exploration tasks.

For the challenges addressed by Enerzyme, we aim to modularly extend capabilities for charge prediction and electrostatic interactions (e.g., as implemented in PhysNet[129] and SpookyNet[130]) to a broader range of GNN-based NNP architectures. We also adapt them specifically to QM clusters embedded in implicit solvent environments. We note that most modern GNN-based NNPs, despite architectural differences, follow a common framework that consists of elemental and geometric embeddings, message-passing interactions, and atomic energy readouts. Based on this observation, we developed Enerzyme with a modular NNP design mode (Supporting Information Figure S3). In Enerzyme, the core message-passing interactions are isolated as a component that maps molecular graph embeddings to atomic features. Pre-core layers encode and aggregate system-level information into the embedding space with the interface beyond elemental and geometric descriptors, while post-core layers decode atomic features into physical quantities, apply corrections or physical priors, and compute derived quantities. This modular design allows different NNP architectures to be extended beyond their original formulations, combining components from multiple models and enabling flexible construction of NNPs tailored to specific simulation tasks. In this work, we specifically exploit this framework to equip models with electrostatics adapted to QM clusters in implicit solvent environments (Figure 2 and Supporting Information Table S11).

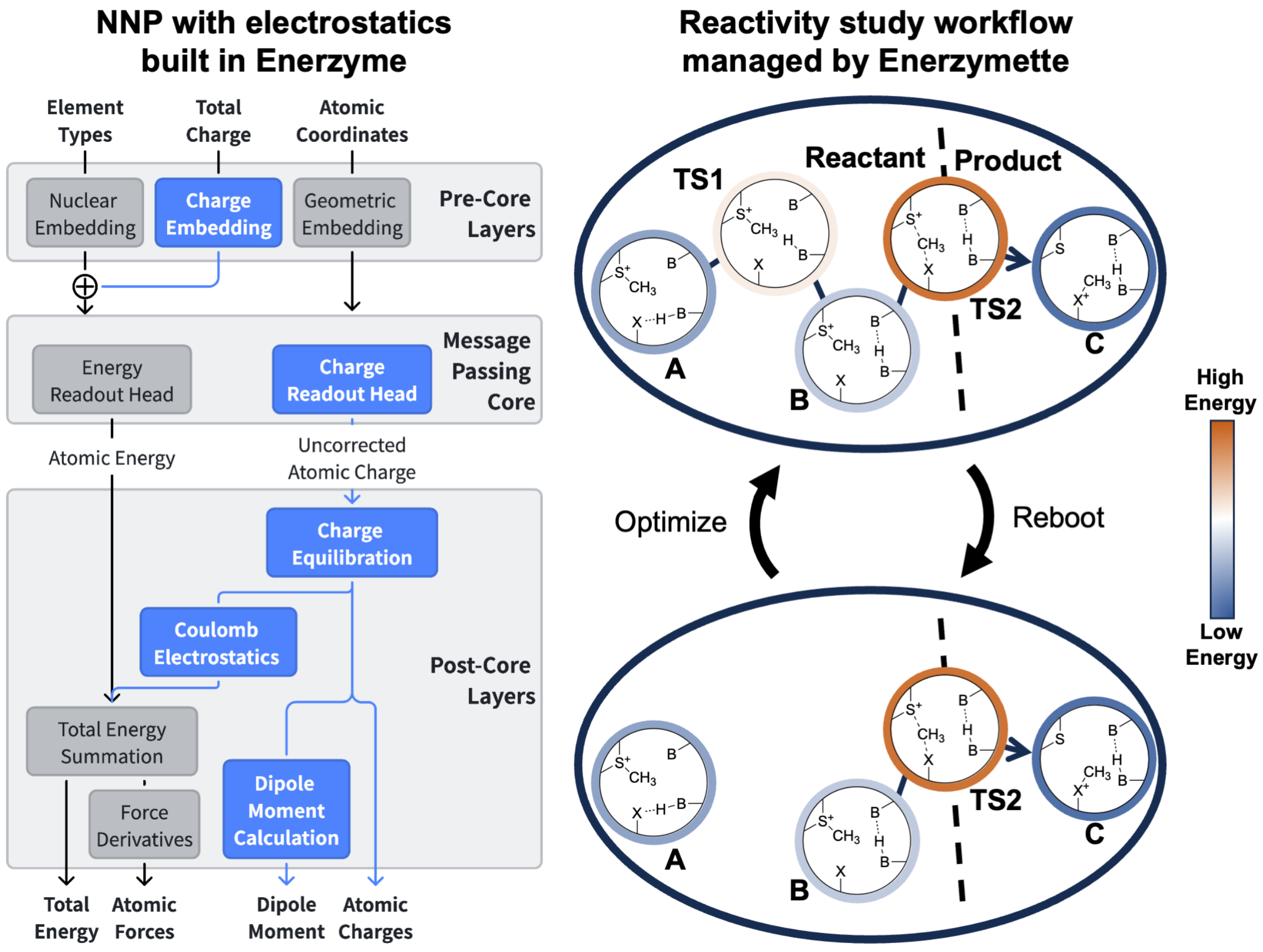

**Figure 2.** An overview of studying reactivity in enzyme cluster models with electrostatics-aware NNPs. Left: We implemented NNPs in a modularized way in our Enerzyme package, where the key modules that equip an NNP with electrostatics are shown in blue. Right: We perform flexible scans or NEBs to study reactivity and evaluate NNPs using our Enerzymette workflow manager. A methyl-transfer reaction from an SAM sulfonium group to a nucleophilic atom X with hydrogen bond partners B is used to illustrate the reaction path. The workflow starts from a reactant conformation (**A**) toward a product conformation (**C**). Once it finds an intermediate structure (**B**) as a local minimum inside the path, it cuts the path into segments at the local minimum and compares the current estimate of TS energies for both segments. It drops the segment that contains the TS of a lower energy (**TS1**), keeps the segment that contains the TS of a higher energy (**TS2**), and reboots the simulation based on the kept segment. It tracks the rate-determining barrier on the complex PES of enzyme clusters by such iterative optimizations until convergence to an elementary reaction.

First, a total-charge embedding is introduced in the pre-core layers to distinguish between clusters with different total charges. After the message-passing phase, atomic features are decoded not only into atomic energies but also into atomic charges in the post-core layers, enabling the model to predict charge distributions as in electronic structure analysis. To ensure consistency with the total charge, $Q$, the raw atomic charge predictions $\tilde{q}_i$ are corrected as $q_i$ through a charge-conservation post-core layer with a linear rescaling (equation 2):

$$q_i = \tilde{q}_i - \frac{1}{N}\left(\sum_i \tilde{q}_i - Q\right). \qquad (2)$$

The corrected atomic charges are then coupled back to the PES through a Coulomb potential post-core layer as a long-range electrostatic energy term (equation 3):

$$E_{\text{ele}} = \frac{1}{4\pi\epsilon_0\epsilon}\sum_{i<j}\frac{q_i q_j}{\chi(r_{ij})}, \qquad (3)$$

where $\epsilon_0$ is the vacuum permittivity, $\epsilon$ is the relative dielectric constant, $r_{ij}$ is the distance between atom $i$ and $j$, and $\chi$ is a damping function. Importantly, the electrostatic layer adopts the same dielectric screening $\epsilon = 10$ as the implicit solvent model used in the QM cluster calculations, ensuring consistent long-range electrostatic behavior between the NNP and reference PES. A dipole calculation layer is additionally included in post-core layers to compute the molecular dipole moment $\boldsymbol{\mu}$ from the predicted atomic charges and atomic positions $\boldsymbol{r}_i$ (equation 4):

$$\boldsymbol{\mu} = \sum_i q_i \boldsymbol{r}_i. \quad (4)$$

enabling multi-objective supervision (i.e., simultaneously training using information from) atomic charges on direct labels, electrostatic observables, and PES properties. The overall loss function is written as a weighted sum of loss terms over all predicted quantities (equation 5):

$$\ell = w_E \ell_E + w_{\boldsymbol{F}} \ell_{\boldsymbol{F}} + w_{\boldsymbol{\mu}} \ell_{\boldsymbol{\mu}} + w_Q \ell_Q + w_q \ell_q, \qquad (5)$$

where every regression loss term $\ell_x$ is weighted by $w_x$, applying to each component of the property $x$ between its predicted value and the corresponding label in the training set, including total energy ($E$), atomic forces ($\boldsymbol{F}$), dipole moment ($\boldsymbol{\mu}$), uncorrected total charge ($Q$), and corrected atomic charges ($q$). For the atomic charge labels, we chose the 1.2×CM5 scheme because they are also compatible with molecular mechanics electrostatics[113]. Thus, learning this charge scheme provides both physically grounded descriptors of charge redistribution along the reaction coordinate and a model of long-range electrostatic interactions as a component of the PES. This framework adapts the electrostatic architectures originally developed in PhysNet and SpookyNet to implicit solvent dielectric environments, while also enabling models without native electrostatics, like MACE[131], to be equipped with the same electrostatic treatment.

For the challenges addressed by Enerzymette, the high dimensionality and ruggedness of enzyme-cluster PESs make identifying key reaction steps and transition states (TS) highly challenging, requiring development of more specialized, automated workflows. Flexible scans or NEB calculations frequently identify additional intermediates between the reactant and the product, such as hydrogen-bond network rearrangements and residue conformational changes, requiring extensive manual inspection and iterative restarting of simulations. Such manual intervention not only lessens the speed advantage of NNPs but also makes rigorous comparisons between DFT and NNP simulations inefficient and intractable. During flexible scans or NEB calculations automated by Enerzymette, when an intermediate local minimum is detected along the path, Enerzymette terminates the current optimization, records the intermediate as a newly found reactant or a product based on its relative position to the barrier within the reaction pathway. The code automatically reboots a new scan or NEB optimization on the side with the higher barrier until a single elementary reaction is obtained (Figure 2, Supporting Information Figures S4 and S5). This iterative workflow effectively tracks the rate-determining TS (i.e., the highest-energy barrier) while recording the explored local minima in reactant and product regions.

We note that on such high-dimensional PESs, the multistep optimization, stopping, and restarting procedures in iterative NEBs become extremely sensitive to the optimizer and the workflow. To isolate the influence of the PES itself from these simulation factors, we align the optimizer and workflow manager between DFT and NNP simulations as closely as possible. Specifically, Enerzymette interfaces ORCA as an external optimizer with both TeraChem and Enerzyme. In this framework, DFT in TeraChem and the NNP in Enerzyme only provide the energies and forces required by the optimizer, while the optimizer generates initial guesses, updates structures, and notifies the workflow manager of intermediate detections to terminate the current run (Supporting Information Figure S6). This design minimizes differences beyond the PES itself.

Furthermore, even if an NNP reproduces the DFT PES locally, small deviations may significantly affect the subsequent optimization trajectories. Instead, we focus on whether the NNP maintains the key mechanistic observables relevant to enzymatic reaction studies. Based on the intermediates and approximate TS identified from iterative scans or NEB calculations, we compute the overall energy span and reaction energy change as the primary energetic metrics. After an iterative scan or NEB converges, the energy difference between the lowest-energy product and lowest-energy reactant is defined as the reaction energy, while the energy difference between the estimated rate-determining transition state and the lowest-energy reactant is defined as the energy span of the reaction. These quantities characterize the kinetic and thermodynamic behavior of the reaction while reducing sensitivity to minor PES differences. For structural metrics, we examine the geometry of the reaction center in the $S_N2$ methyl transfer mechanism, focusing on the attacking angle and the forming and breaking bond distances that characterize the TS (Figure 3). Besides the PES, we also evaluate whether our NNP-predicted atomic charges capture the charge redistribution between reactants, polarization of surrounding residues along the reaction coordinate, and the relationship between substrate charge and reactivity. We consider these metrics to representatively reflect the practical requirements of structure–reactivity studies in enzyme QM cluster models.

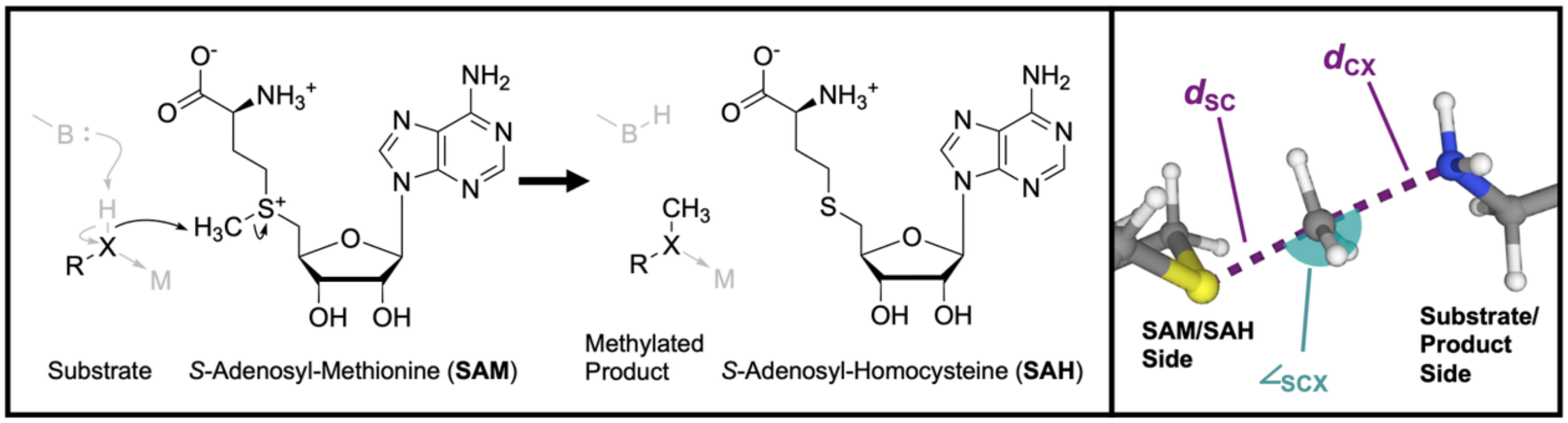

**Figure 3.** The reaction scheme in SAM-MT enzymes and the reaction center geometry of the transition state. Left: The SAM-MT has an *S*-Adenosyl-Methionine (SAM) cofactor as the methyl group donor that transfers the methyl group from its sulfonium group to the nucleophilic atom X of the substrate. The X is typically O, N, C, or S, and may be deprotonated by a base (denoted by B in the scheme) and coordinate with a metal atom (denoted by M in the scheme). Right: We use 3 local geometry descriptors to characterize the transition state in our evaluation of the NNPs. $d_{SC}$ is the distance between the SAM's sulfonium sulfur atom S (in yellow) and the methyl carbon atom C (in gray), and $d_{CX}$ is the distance between the methyl carbon atom C and the nucleophilic atom X of the substrate (a nitrogen in blue as an example here). $\angle_{SCX}$ is the $S_N2$ attacking angle between the S, C, and X atoms.

**3c. Benchmarking neural network PESs of enzyme cluster models in mechanistic simulations.**

We first evaluated Enerzyme-SpookyNet and Enerzyme-MACE NNPs trained on a specific QM cluster dataset and tested on the same system. Here we introduced two MTases that were previously investigated in our mechanistic QM/MM study[64]: PfPMT, the phosphoethanolamine methyltransferase in *Plasmodium falciparum*[139], and HcgC, a gene product of the *hcgC* gene cluster responsible for the methyl transfer in the iron guanylylpyridinol cofactor biosynthesis[86]. After extracting QM cluster models from the QM regions in the QM/MM study, the PfPMT system features an *N*-methyltransferase with a 284-atom cluster, while the HcgC system features a *C*-methyltransferase with a 545-atom cluster. Therefore, this test evaluates the accuracy of Enerzyme methods on distinct MTase nucleophile identities and substantially different system sizes. During iterative NEB simulations, both NNPs remain stable throughout the optimization process and successfully converge to chemically reasonable reaction pathways. On the 284-atom PfPMT cluster, both models reproduce the DFT reaction energy and kinetic barrier within 3.5 kcal/mol, while Enerzyme-SpookyNet achieves a kinetic barrier error below 0.5 kcal/mol. When extended to the larger 545-atom HcgC cluster, both models maintain stable simulations and predict the reaction kinetic barrier within 2.5 kcal/mol of the DFT reference, but Enerzyme-MACE shows a notably higher error on the reaction energy of 5.9 kcal/mol (Figure 4 and Supporting Information Table S12).

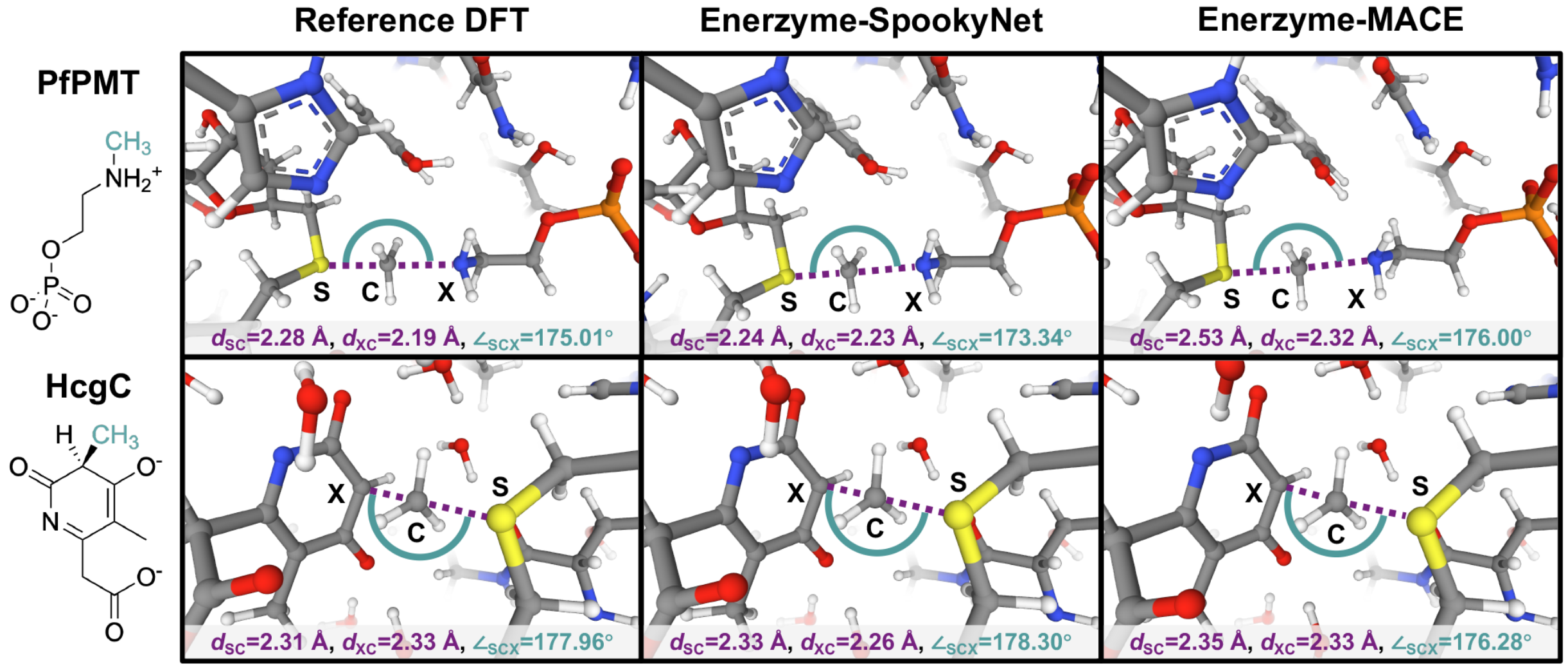

**Figure 4.** NEB-Estimated TS structures of the reaction center of methyl group transfer in HcgC and PfPMT cluster models by the reference DFT and Enerzyme-NNPs. The chemical structures of the methylated substrates in both enzymes are shown on the left with the transferred methyl group in green. $d_{SC}$, $d_{CX}$, and $\angle_{SCX}$ are the local geometry descriptors defined in **Figure 3**. Color scheme: carbon in gray, hydrogen in white, oxygen in red, nitrogen in blue, sulfur in yellow, and phosphorus in orange.

We next extended this system-specific training and evaluation to five catechol *O*-methyltransferase (COMT)[140] cluster models that are around 300 atoms in size, which we constructed from PDB structures using QuantumPDB. Although these clusters all have the same enzyme identity, the protein-ligand complexes span chemically diverse ligands ranging from near-native substrates to inactive inhibitors (Scheme 1). Comparing NNP flexible scans on the same individual structures used to generate the DFT results, Enerzyme-SpookyNet achieves an average error of 1.6 kcal/mol on kinetic barriers and 4.4 kcal/mol on reaction energies. Enerzyme-MACE achieves an average kinetic barrier error of 1.4 kcal/mol and an average reaction energy error of 2.9 kcal/mol (Supporting Information Figure S7). These errors correspond to both over- and underestimates with no clear bias from the NNPs. These results demonstrate the robustness of combining automated QM cluster construction with QuantumPDB and NNP training within the Enerzyme framework for enzyme cluster simulations.

**Scheme 1**. Ligand structures and their associated PDB IDs for the protein-ligand complex crystal structures of the 5 COMT systems used in the Enerzyme-NNP system-specific evaluations. The nucleophilic atoms in the methyl transfer reactions are colored.

| Ligand | | | | | |
|---|---|---|---|---|---|
| PDB ID | 2ZVJ | 4XUD | 4XUE | 6LFE | 7XJB |

We also examined whether currently available universal NNPs for finite-size molecular systems, including UMA[102] and MACE-mh[141], can directly replace DFT calculations in these QM cluster models without system-specific training[142]. However, single-point energy evaluations on DFT flexible scan trajectories show that these models tend to significantly overestimate the reaction barriers compared to system-specific models trained in Enerzyme (Figure 5). This behavior suggests that universal NNPs trained on vacuum PESs cannot be directly transferred to QM cluster models under implicit solvent conditions, highlighting the necessity of targeted retraining on system-specific electronic structure data.

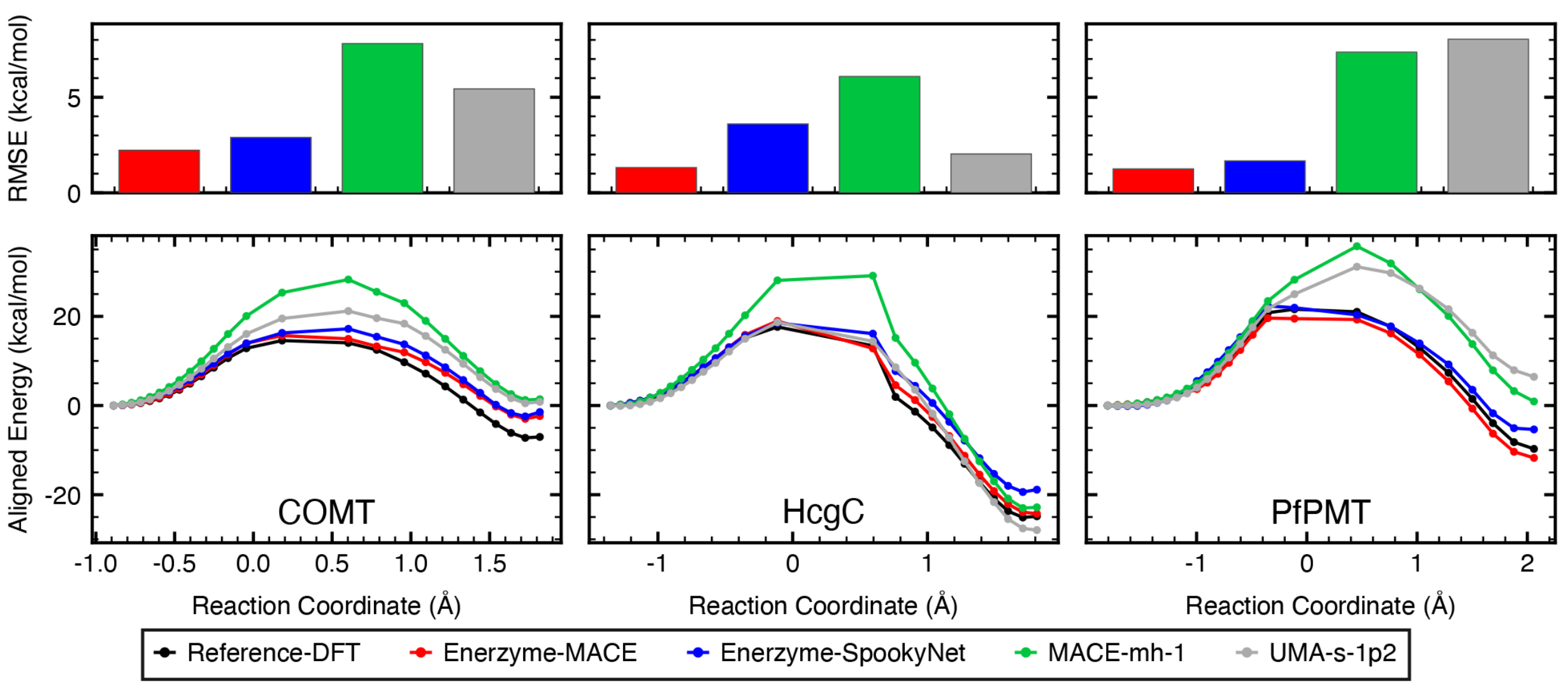

**Figure 5.** Single-point energy evaluation of NNPs along DFT flexible scan trajectories in COMT, HcgC, and PfPMT cluster models. The COMT cluster comes from PDB ID 2ZVJ. The initial energies of different energy curves are aligned to be the same, and the root mean square errors (RMSE) are calculated between the energy curves of each NNP and the reference DFT after energy alignment. The reaction coordinate is the difference between the breaking bond length and the forming bond length $d_{CX} - d_{SC}$ (defined in Figure 3).

Beyond energetics, correct characterization of transition-state structures is critical for mechanistic studies. We find that the models can accurately reproduce the local geometries of DFT transition states in NEB simulations (Figure 4). In the PfPMT system, Enerzyme-SpookyNet reproduces the forming and breaking bond lengths at the transition state within 0.05 Å, while Enerzyme-MACE exhibits larger deviations in bond lengths. Nevertheless, both models predict the nucleophilic attack angle within 2° of the DFT reference. The larger structural deviation of Enerzyme-MACE is caused by a prediction of a too-late transition state (i.e., the leaving methyl group is too close to the amino nitrogen nucleophile) despite good prediction of the angle in the TS. In the larger HcgC system, both models achieve bond-length errors below 0.04 Å and maintain attack-angle deviations within 2°, consistent with the expected $S_N2$ TS geometry formed in the enzyme environment.

Despite the thousands of degrees of freedom involved in these QM cluster models, the global structures obtained from NNP and DFT geometry optimizations and NEB calculations remain highly consistent when starting from the same initial structures. For the PfPMT cluster, the root mean square deviation (RMSD) of all unconstrained atoms between NNP- and DFT-optimized reactant, product, and TS structures is around 0.5 Å. For the larger 545-atom HcgC cluster, the agreement is even stronger, with RMSDs of only 0.2–0.4 Å (Supporting Information Table S13). These results indicate that Enerzyme-NNPs can learn high-quality PES landscapes for enzyme cluster models from fewer than 1000 steered-MD configurations and drive stable and reliable reaction-path exploration in large SAM-MT clusters.

To test whether the stability of NNP-driven NEB simulations in enzyme cluster models imposes stricter requirements on the local details and global landscape of the PES than simpler tasks such as single-point energy prediction or geometry optimization, we additionally trained a set of Enerzyme-PhysNet models with a smaller capacity (i.e., fewer trainable parameters) and a simpler architecture. Although these models exhibit satisfactory performance in simpler single point energy evaluation tasks, they fail in more demanding NEB simulations. We first find that single-point energy accuracy of reaction barriers cannot reliably predict NEB stability, even when the evaluation is already performed on DFT flexible scan trajectories. For single-point energy evaluation on a DFT-scan trajectory of the HcgC cluster, Enerzyme-PhysNet shows an RMSE of 4.3 kcal/mol using the initial reactant energy to align the PESs, which is very similar to 3.6 kcal/mol from Enerzyme-SpookyNet, yet fails to converge to a correct TS structure during NEB (Supporting Information Figure S8). In addition, although both flexible scans and NEB provide an assessment of NNP fidelity along reaction pathways, scan accuracy alone still cannot reflect NEB reliability. In the PfPMT system, Enerzyme-PhysNet reproduces the kinetic barrier and reaction energy within approximately 2 kcal/mol in flexible scan simulations but still fails to converge to the correct transition-state structure in NEB calculations (Supporting Information Figure S9). These observations indicate that NEB simulations are more demanding on PES quality, which may not be revealed by single-point metrics or simpler scanning procedures. Therefore, evaluating NNPs in realistic NEB workflows is essential for assessing their reliability in practical reaction mechanism studies.

### 3d. The role of electrostatics in Enerzyme-NNPs for enzyme cluster models.

Electrostatic interactions play a central role in enzyme catalysis. In SAM-dependent methyltransferases, the SAM cofactor itself is highly polarized because its charge is concentrated across the sulfonium center, carboxylate, and amino moieties. Together with highly charged substrate motifs such as phosphates in PfPMT or $Mg^{2+}$-coordinated phenols in COMTs, this leads to complex charge distributions and long-range electrostatic interactions that strongly influence the reaction. Therefore, we studied how the electrostatic components in Enerzyme-NNPs influence their accuracy and interpretability in MTase cluster models.

First, we find that direct supervision of atomic charges (i.e., including atomic charges in the loss function) is essential for structure-reactivity simulations in enzyme clusters (Figure 6). In both the PfPMT and HcgC systems, reducing the atomic-charge supervision weight by two orders of magnitude (Qa1 in Figure 6) significantly increases the NEB energy and structural errors. In several cases, the large structural errors are so severe that the models fail to converge to a reasonable TS structure. These observations indicate that direct supervision on electronic charge distributions is not merely aligning the model outputs to electronic structure analysis but is shaping the detail of the reactive PES of highly polarized enzyme cluster environments for stable and chemically meaningful simulations.

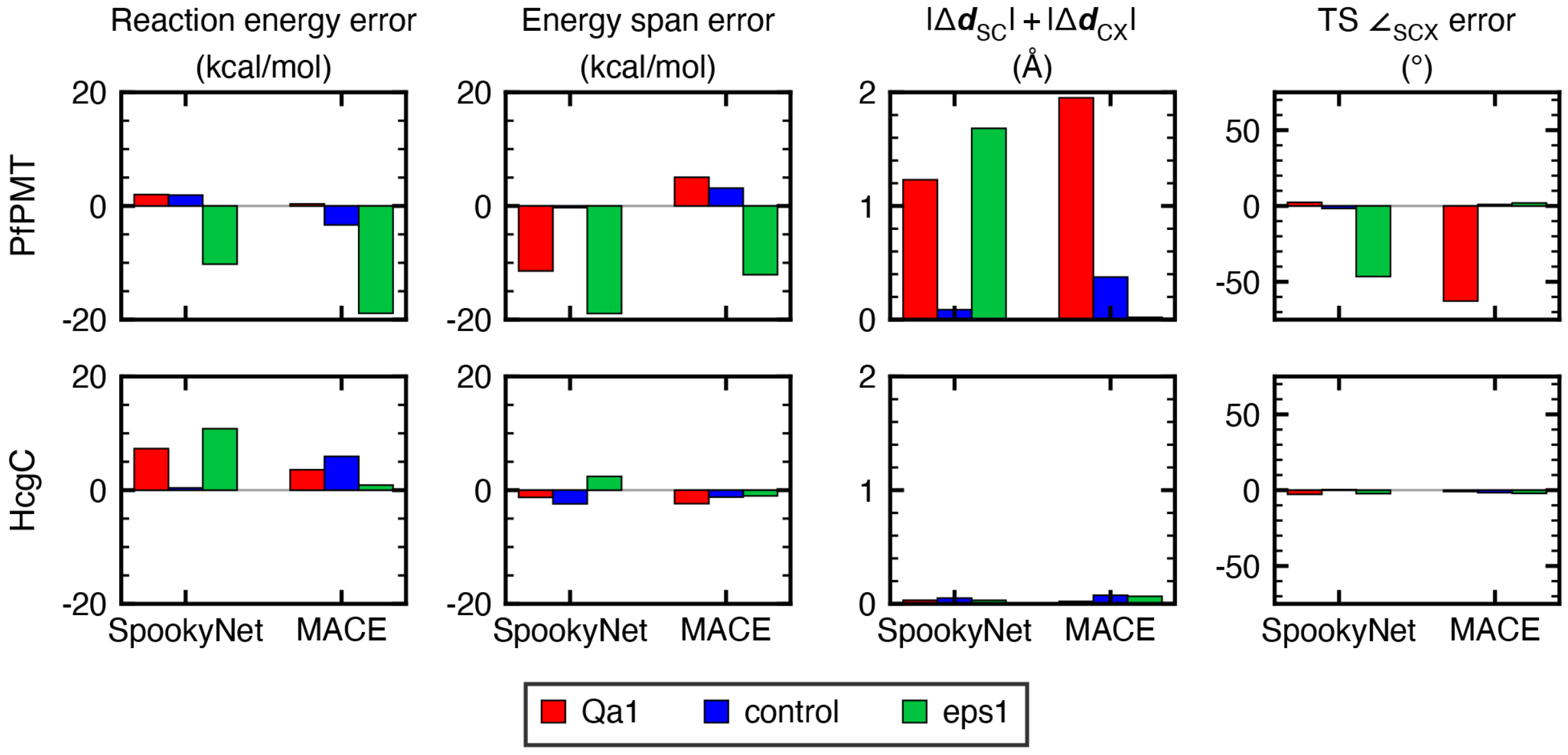

**Figure 6.** NEB performance for SpookyNet and MACE trained on PfPMT and HcgC enzyme cluster structures with different electrostatic settings in Enerzyme. "Qa1" stands for weak supervision on atomic charges, using a loss weight of 1 instead of 100 in the control (i.e., default) NNP. "eps1" stands for using a dielectric constant of 1 instead of 10 in the control NNP models. The energy change error is the error of the energy difference between the lowest-energy product structure and the lowest-energy reactant structure found in the NEB. The energy span error is the error of the energy difference between the highest-energy barrier and the lowest-energy reactant structure found in the NEB. $\Delta d_{SC}$ and $\Delta d_{CX}$ denote the errors in the bond-length metrics $d_{SC}$ and $d_{CX}$ (defined in Figure 3) of the TS structure estimated by CI-NEB, respectively. The TS $\angle_{SCX}$ error is the error of the attacking angle metric (defined in Figure 3) of the TS structure estimated by CI-NEB. All errors are calculated using the DFT results at the same level of theory as the NNP training data with the same CI-NEB protocol as the ground-truth reference.

Second, in our QM cluster calculations, the systems are embedded in implicit solvent environments rather than vacuum, and the electrostatic interactions are therefore screened by a finite dielectric constant representing the protein environment. We find that using an inconsistent dielectric constant substantially degrades the quality of NEB simulations (Figure 6). In both the PfPMT and HcgC systems, using no dielectric screening (i.e., vacuum labeled eps1 in Figure 6) also leads to significantly larger errors in reaction energetics and TS geometries. These observations demonstrate that long-range electrostatic interactions in enzyme clusters cannot be treated independently from the dielectric environment in which the reference electronic structure calculations are performed. When coupling charge distribution with PES, incorporating physically consistent electrostatic priors into the NNP architecture is therefore essential for obtaining contributive electrostatic components in the reactive PES.

The atomic charges as coordinate-dependent functions learned together with the PES serve as dynamic descriptors of electronic structure evolution along the reaction pathway. In our multi-objective supervision, the predicted atomic charges are also guided by the dipole objective and coupled with the potential energy surface, in addition to the supervision on the reference atomic charge labels. Thus, we evaluate their validity through their general trends and correlation with the reference, DFT-derived atomic charge (i.e., 1.2×CM5) trends on the same reaction trajectories instead of the accuracy of reproducing the reference charge scheme.

We examined the charge transfer within the reaction center, including the SAM cofactor and substrate, and also evaluated the role of surrounding residues. Consistent with our previous analysis in QM/MM studies, we observe substantial positive charge transfer from the SAM cofactor to the substrate during methyl transfer, while significant charge separation is preserved near the TS region (Figure 7). Specifically, in the PfPMT system, the model with a reduced atomic-charge supervision weight (Qa1in Figure 7) exhibits too-conservative charge predictions, although models employing incorrect dielectric constants (eps1 in Figure 7) still qualitatively reproduce these charge-transfer and polarization trends. This latter model underestimates both the degree of charge transfer between the cofactor and substrate and the extent of charge separation. In the PfPMT system, the NNP predictions with full charge supervision also reproduce trends in charges of nearby amino acid residues with strong correlations to the reference DFT charge variations (Figure 7). However, when the atomic-charge supervision weight is reduced (Qa1 in Figure 7), the correlation for several residues, including TYR9 and HID122, degrades significantly. These observations suggest that direct supervision of atomic charges improves the model's ability to resolve fine-grained electronic structure variations inside complex enzyme clusters.

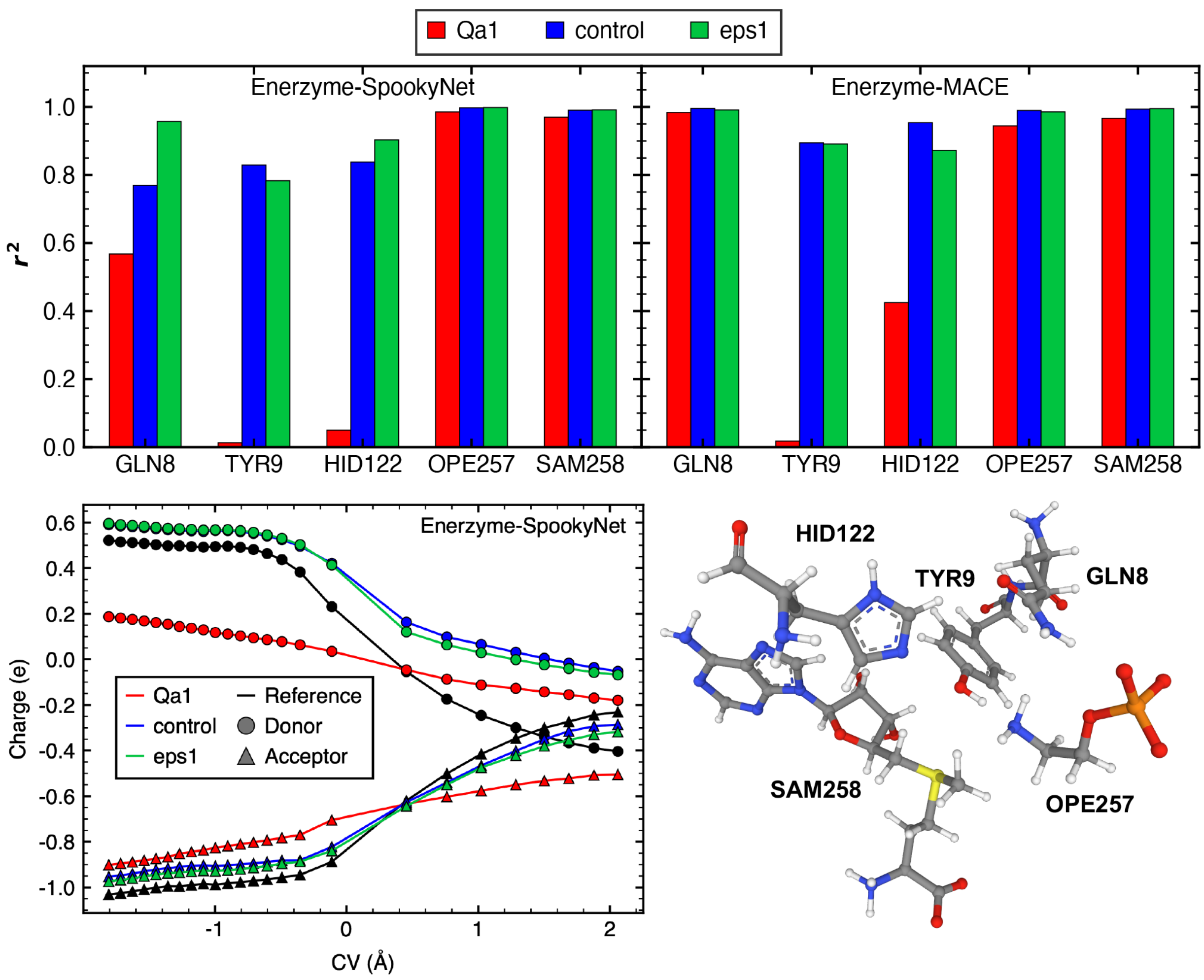

**Figure 7**. The analysis of charge-transfer and polarization effects from NNP-predicted atom charges. Top: the correlation between selected residue charges (summed atomic charges) predicted by the Enerzyme-NNPs and the one obtained from electronic structure analysis during single-point predictions along the reference-DFT flexible scan trajectory of the PfPMT cluster. The $y$-axis is Pearson's $r^2$ coefficient, and the $x$-axis is grouped by the residue names in and around the reaction center. Both Enerzyme-SpookyNet and Enerzyme-MACE models are tested with different electrostatic settings in Enerzyme. "Qa1" stands for weak supervision on atomic charges, using a loss weight of 1 instead of 100 in the control NNP models. "eps1" stands for using a dielectric constant of 1 instead of 10 in the control NNP models. Bottom left: the charge-transfer analysis of the methyl transfer between SAM258 and OPE257 on the same reference-DFT flexible scan trajectory (See Computational details). The Qa1, control, and eps1 NNP models follow the same naming convention, and the reference curve stands for the 1.2×CM5 charges. Bottom right: an illustration of the selected residues in and around the PfPMT reaction center with their residue names and reactant-like geometry. Color scheme: carbon in gray, hydrogen in white, oxygen in red, nitrogen in blue, sulfur in yellow, and phosphorus in orange.

As electronic structure descriptors, the NNP-predicted atomic charges also reflect chemically meaningful reactivity trends in methyl transfer reactions. We compared five COMT cluster systems with chemically diverse substrates (Scheme 1). The predicted charge on the nucleophilic atom reflects the substituent effects of functional groups on the aromatic rings and hetero atoms and exhibits strong correlations with both the reaction barriers and reaction energy changes obtained from the same NNP-driven scans for each of these substrates. Interestingly, these correlations are more pronounced than those obtained from conventional DFT-derived atomic charge schemes, even including the reference 1.2×CM5 charges used during training (Figure 8 and Supporting Information Figure S10). Some charge-fitting procedures, such as RESP, prioritize reproducing the global electrostatic potential around the molecular surface, which can ignore detailed charge concentration features buried deep inside enzyme clusters. In contrast, our Enerzyme-NNP is simultaneously constrained by global observables, local atomic supervision, and PES consistency during training. This multi-level supervision enables learning a charge representation that captures subtle electronic effects associated with reactivity, suggesting that the NNP develops a chemically interpretable electronic descriptor based on a predefined atomic charge scheme (i.e. 1.2×CM5) and maintains or even surpasses its predictive power on reactivity.

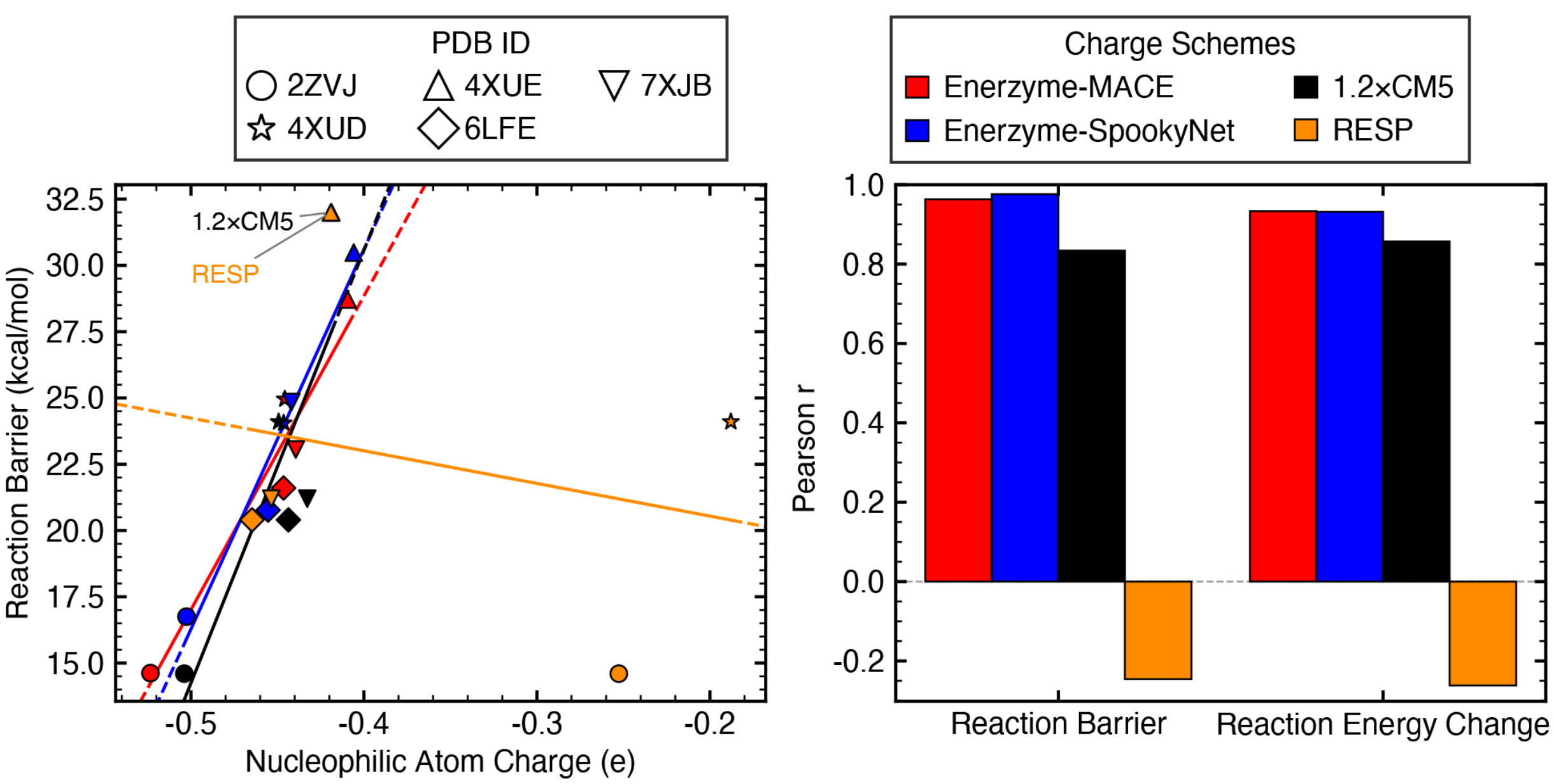

**Figure 8.** The correlation between the partial charge of the nucleophilic atom of COMT substrates and the reaction barrier and the reaction energy change. For all systems from 5 PDB IDs (Scheme 1), their reaction barriers and atomic charges obtained from reference DFT flexible scans and an Enerzyme NNP flexible scan with atomic charge prediction are shown with a legend corresponding to each system at the top left. The four charge schemes are colored according to the legend at the top right: Enerzyme-MACE (red), Enerzyme-SpookyNet (blue), 1.2xCM5 (black), and RESP (orange). Left: The lines show the linear regression between the predicted reaction barrier of the 5 systems and the predicted nucleophilic atom charge with corresponding markers for different PDB IDs. Right: The Pearson *r* coefficients for the linear regression between the predicted reaction barrier/reaction energy change for the 5 systems and the predicted nucleophilic atom charge across different charge schemes.

### 3e. Evaluating generalizability and computational cost of Enerzyme-NNPs.

To amortize the cost of quantum chemistry dataset labeling and model training, it is desirable that a single NNP achieves reliable accuracy across multiple enzyme-substrate systems.

Even if across the same enzyme family, like MTases, this requires generalization over both perturbations in active site protein environments and the broad chemical diversity of substrates. To assess this transferability, we performed cross-evaluation on various COMT cluster models with distinct structure sources and different substrate identities.

We first investigate the zero-shot transferability of a model trained on a specific COMT cluster dataset to the iterative flexible scan simulation of other COMT clusters. We find that the transferability of both Enerzyme-SpookyNet and Enerzyme-MACE remains strongly dependent on substrate chemistry and reaction energetics. In particular, models trained on other systems perform substantially worse on the 2ZVJ complex than models trained directly on 2ZVJ itself. For example, compared to the Enerzyme-SpookyNet model trained on 2ZVJ, which yields only +2.2 kcal/mol energy-span error and +1.7 kcal/mol reaction-energy error, models trained on 4XUD, 4XUE, and 7XJB significantly overestimate the reaction barrier of 2ZVJ by +7.1, +7.9, and +9.9 kcal/mol, respectively, while the reaction-energy errors further increase to +9.3, +14.8, and +15.5 kcal/mol (Supporting Information Figure S7). Enerzyme-MACE exhibits similar behavior.

One possible explanation is that the models become specialized toward the characteristic functional groups in the ligand present in their own training set, including the boronic acid in 4XUD, the pyridone in 4XUE, and the chlorine, nitro group, and oxadiazole motifs in 7XJB, but they fail to generalize to the distinct pyranone-like scaffold of 2ZVJ. In addition, 2ZVJ has the lowest DFT barrier and most exergonic reaction energy among the five systems, suggesting that models trained on the other systems may partially overfit their respective energetic trends and therefore systematically overestimate the TS and product energies, rather than learning a generalized description of how phenolic nucleophile reactivity depends on chemical environment.

In contrast, models trained on the 6LFE system exhibit substantially better transferability to 2ZVJ. For example, the Enerzyme-SpookyNet and Enerzyme-MACE models trained on 6LFE produce only +0.1 and +2.7 kcal/mol barrier errors on 2ZVJ, respectively. This improved transferability likely originates from the greater chemical similarity between the 6LFE and 2ZVJ ligands, as well as their similar reaction-energy trends. Overall, Enerzyme-MACE demonstrates stronger zero-shot generalization than Enerzyme-SpookyNet. Averaged over the four models trained on other systems and evaluated on 2ZVJ, Enerzyme-MACE achieves mean barrier and reaction-energy errors of 5.9 and 10.0 kcal/mol, slightly better than the corresponding 6.2 and 11.1 kcal/mol errors of Enerzyme-SpookyNet. The difference becomes more pronounced when transferring to the 4XUE system, where Enerzyme-MACE reaches only 2.3 kcal/mol average barrier error and 2.8 kcal/mol average reaction-energy error, significantly outperforming Enerzyme-SpookyNet with corresponding errors of 5.8 and 8.9 kcal/mol. Interestingly, the Enerzyme-MACE model trained on the nitrogen-free 2ZVJ system still generalizes reasonably to the nitrogen-containing heterocycle of 4XUE, with a barrier error of only +0.80 kcal/mol. This behavior suggests that Enerzyme-MACE may partially transfer chemically relevant information from similar but common motifs in amino acids, such as the tryptophan residue, whereas Enerzyme-SpookyNet fails to generalize similarly and instead underestimates the 4XUE barrier by 9.17 kcal/mol.

Overall, these observations suggest that models trained on single enzyme–substrate systems are insufficient for a reliable application across the broad chemical space of small-molecule ligands. We therefore further investigate whether NNPs can extract transferable chemical knowledge from combined datasets of multiple enzyme–substrate systems. To this end, we combined the datasets from five COMT systems (Scheme 1) and trained models using a subset that includes 200 structures per system with a training/validation ratio of 7:1, which in total has the same size (1000 structures) as the system-specific training. We also experimented with 2× and 4× data ratios, which correspond to 400 and 800 structures per system, respectively. In addition to the performance on the five systems present in the combined dataset, we further benchmark the resulting models on four additional COMT systems containing ligands completely absent from the training set (Supporting Information Figure S11). Those models show promising accuracy in the reaction barrier estimation with iterative flexible scans, and their in-dataset performance can be systematically improved by scaling up the dataset. Trained on the 4× datasets, Enerzyme-SpookyNet reaches 1.7 kcal/mol average barrier error on the in-dataset systems, already approaching the previously obtained 1.6 kcal/mol error achieved by separately trained, single-system models using 875 structures for each system. Furthermore, using only 175 structures per system, corresponding to the same total dataset size as the original single-system training sets, Enerzyme-MACE achieves 2.2 kcal/mol average in-dataset barrier error, which is comparable to the 1.4 kcal/mol error obtained from individually trained models. Enerzyme-MACE consistently exhibits higher overall data efficiency than Enerzyme-SpookyNet, as it achieves a 2 kcal/mol out-of-dataset error with the 1× data ratio, whereas Enerzyme-SpookyNet requires the 4× data ratio and 1.5× more total cost (Figure 9 and Supporting Information Table S14). Among all the models, out-of-dataset errors remain comparable to in-dataset errors, indicating that they indeed learn generalized chemical knowledge about reactivity trends within the COMT ligand chemical space. We also observe that, for Enerzyme-MACE trained on the largest dataset, the out-of-dataset accuracy becomes slightly worse than that of those trained on smaller datasets, warning that it may already begin to overfit the training systems.

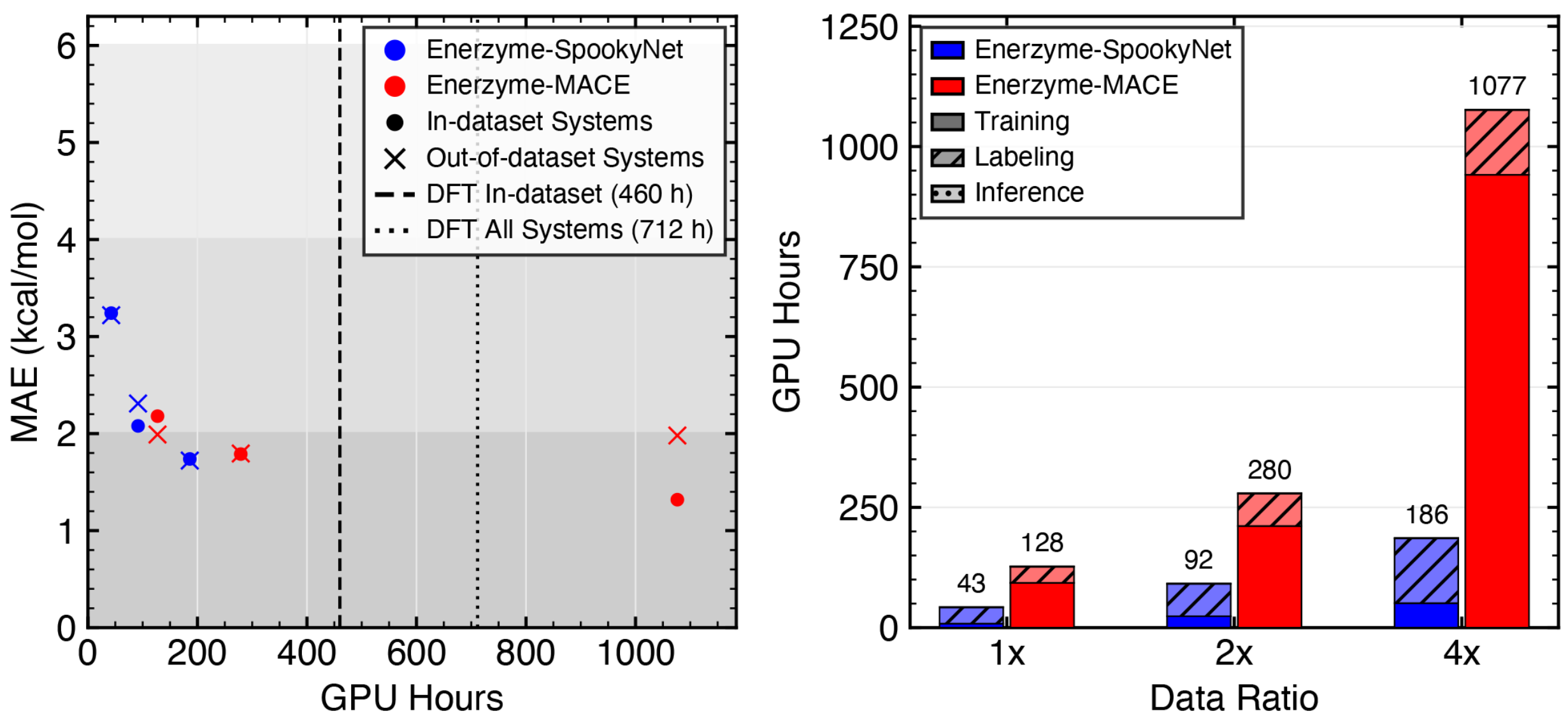

**Figure 9.** Accuracy–cost trade-offs of Enerzyme-NNPs in iterative scan simulations on COMT cluster models. Left: The *x*-axis reports the total GPU hours of a complete Enerzyme-NNP workflow, while the *y*-axis reports the mean absolute error (MAE) of reaction energy barriers relative to the reference DFT calculations. Dashed and dotted vertical lines indicate the total GPU time cost of the corresponding iterative scans directly with the reference DFT calculations. Right: Breakdown of the total GPU hours of the complete Enerzyme-NNP workflow into dataset labeling, model training, and iterative scan inference at different dataset scales.

As we scaled up the dataset size, we also analyzed the cost of the full workflow on a single NVIDIA V100 32GB GPU, including dataset labeling, model training, and iterative scan simulations at different dataset sizes, and compared it with the DFT cost with the same computational units. The iterative scan inference cost of 9 systems in total is no more than 1 hour for all models, highlighting that NNPs enable the acceleration of DFT simulation by 2–3 orders of magnitude, which would have required 712 hours in total. When taking labeling and training costs into account, for relatively lightweight models such as Enerzyme-SpookyNet, quantum chemistry labeling dominates the computational (78% with 1× data ratio, Figure 9). In contrast, for larger-capacity architectures such as Enerzyme-MACE, model training itself becomes the major computational bottleneck (73% with 1× data ratio, Figure 9). Due to training instability, we employed highly conservative learning-rate schedules and early-stopping strategies, inflating the training cost of Enerzyme-MACE. As a result, the largest Enerzyme-MACE model becomes more expensive (1077 hours) than directly performing DFT iterative scans on all nine COMT systems (712 hours). Therefore, both the efficiency and accuracy of Enerzyme-MACE still have considerable room for improvement through better training strategies and hyperparameter optimization. But still, both Enerzyme-SpookyNet and Enerzyme-MACE are able to achieve a 2 kcal/mol accuracy in 186 and 127 hours for both in-dataset and out-of-dataset systems, respectively, which are cheaper than the 460-hour DFT simulation on those 9 systems. Our results suggest the potential of scaling a generalizable NNP with reaction datasets of more diverse enzymes and applying it to more unseen systems. While the NNP-driven simulation remains consistently far cheaper than DFT-driven, the DFT labeling and model training cost can be amortized across a larger number of target systems and downstream tasks, making the advantage of NNP increasingly significant.

## 4. Conclusions

In this work, we developed an integrated framework for NNP-driven mechanistic simulations in QM cluster models that we demonstrated on representative MTases. We combined automated QM-cluster construction, reactive dataset generation, electrostatics-aware NNP modules, and iterative reaction-path exploration. These advances enabled NNP development and evaluation for structure–reactivity workflows, as automated through our Enerzyme and Enerzymette software packages, and established a practical route to accelerate QM-level simulations, as demonstrated on MTases. Across the studied MTase cluster models up to 500 atoms, Enerzyme-SpookyNet and Enerzyme-MACE trained on fewer than 1,000 cluster structures for a specific system reproduced DFT reaction energetics and TS structures with good accuracy. We also showed that realistic mechanistic studies impose stricter requirements on PES quality than conventional single-point metrics. Therefore, future NNP development in reactive simulation, such as catalysis and combustion studies, should emphasize these types of evaluations.

Our results further demonstrated that electrostatic information is essential for enzyme-cluster NNPs. Direct supervision of atomic charges and dielectric screening alignment with QM cluster models both substantially improved the stability and accuracy of NEB simulations, while the charge distributions learned under multitask training provided a consistent description of charge transfer and polarization. Furthermore, the predicted atomic charge of the nucleophilic atoms exhibited high correlation with reaction energetics across a wide range of chemistry in COMT enzymes, highlighting its chemically meaningful insights as a reactivity descriptor.

The emergence of transferability across chemically distinct COMT substrates suggests that NNPs begin to learn generalized biochemical reactivity patterns as training datasets are expanded to chemically diverse systems. Extending this framework to additional enzyme families, reaction mechanisms, and substrate chemistry will lead to transferable representations of enzymatic reactivity across broader biochemical space. At the same time, continued advances in NNP architectures, active learning strategies, and the adaptation of NNP foundation models to QM cluster settings may substantially reduce the cost of data labeling and model training required for system-specific applications. With increasingly transferable NNPs and automated workflows, we anticipate that QM-level simulations of enzyme cluster models can be deployed at a scale that is currently intractable with conventional electronic structure methods. More broadly, the strategies developed here for learning and evaluating reactive PES in large-scale, complex biochemical environments may inspire future ML/MM embedding approaches and eventually tractable NNP-driven simulations of full biomolecular complexes in aqueous solutions, enabling more comprehensive and predictive modeling of enzymatic reactivity. Such advances in efficiency may enable large-scale mechanistic studies, mutant screening, and substrate-design campaigns across enzyme families while retaining the interpretability and chemical insight of quantum chemistry.

## ASSOCIATED CONTENT

### Data and code availability

The Enerzyme code with initial QM cluster structures and workflow input files are provided in the Zenodo repository at the DOI https://doi.org/10.5281/zenodo.21018797 and the GitHub repo https://github.com/Benzoin96485/Enerzyme. The Enerzymette code is provided in the Zenodo repository at the DOI https://doi.org/10.5281/zenodo.21018916 and the GitHub repo https://github.com/Benzoin96485/Enerzymette.

**Supporting Information**. Detailed configurations for QM cluster construction, molecular simulations, and NNP training; workflow implementation and algorithms; additional molecular structures, computational results, and data analysis. This material is available free of charge via the Internet at http://pubs.acs.org.

## AUTHOR INFORMATION

### Corresponding Author

*email:hjkulik@mit.edu

### Notes

The authors declare no competing financial interest.

ACKNOWLEDGMENT

This work was supported by the National Institute of General Medical Sciences of the National Institutes of Health under award number R35GM152027 (to H.J.K. and W.L.). Partial support was provided by the MIT Office of Research Computing and Data (ORCD) Seed Fund (to H.J.K. and W.L.) and Molecular Sciences Software Institute (MolSSI) Software Fellowship Program (to W.L.). The authors acknowledge MIT SuperCloud at Lincoln Laboratory Supercomputing Center and MIT Engaging at ORCD for providing HPC and consultation resources that have contributed to the research results reported within this paper.

**For Table of Contents Use Only**

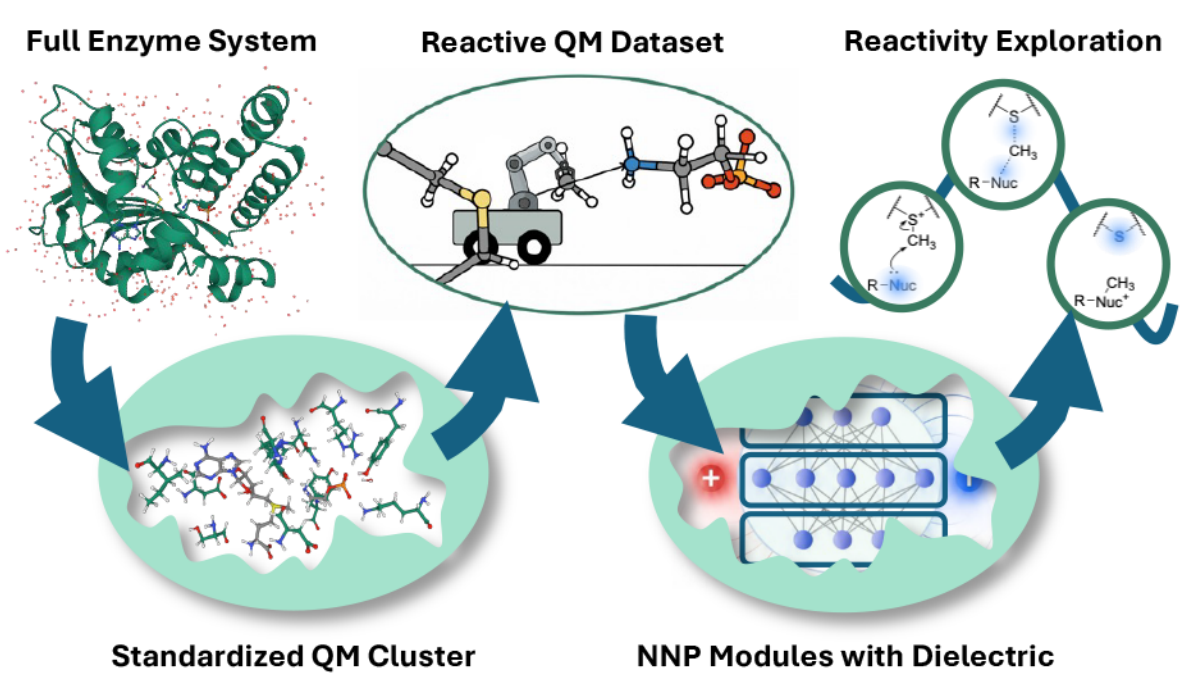

**Supporting Information for**

***Enerzyme: A Framework for Efficient Training of Reactive Neural Network Potentials for Enzyme Catalysis with Application to Methyltransferases***

Weiliang Luo[1,2] and Heather J. Kulik[1,2,*]

[1]*Department of Chemistry, Massachusetts Institute of Technology, Cambridge, MA 02139, USA*

[2]*Department of Chemical Engineering, Massachusetts Institute of Technology, Cambridge, MA 02139, USA*

*corresponding author email: hjkulik@mit.edu

*email: hjkulik@mit.edu

**Contents**

**Table S1.** Structural details of the PfPMT QM cluster model. The *holo* structure of PfPMT used in this study is based on its parent PDB entries 3UJ7 and 3UJB with a peer-reviewed publication[1]. The all-atom structure processed from the crystal structure was downloaded from their supporting information files of our previous QM/MM study on methyltransferases[2], from which the amino acid residues and small molecules in the QM region specified in the supporting information were extracted as the QM cluster model. Cleaved peptide bonds were capped with hydrogen atoms. The table lists the three-letter codes of the residues (Res. Name) included in the QM cluster model, along with their corresponding residue numbers in the reference full-protein structure (Res. Number) and net charges at their determined protonation states. In addition to standard three-letter codes, HID means a neutral HIS where a proton resides at the $N_{\delta}$ of the side chain, OPE is the substrate (2-aminoethyl phosphate), and SAM is the cofactor (*S*-adenosylmethionine). This leads to a QM cluster model with 284 atoms after capping and a net charge of −1.

| Res. Number | Res. Name | Net Charge | Res. Number | Res. Name | Net Charge |
|---|---|---|---|---|---|
| 8 | GLN | 0 | 122 | HID | 0 |
| 9 | TYR | 0 | 165 | TYR | 0 |
| 17 | TYR | 0 | 169 | ARG | +1 |
| 54 | SER | 0 | 237 | LYS | +1 |
| 75 | ASP | −1 | 257 | OPE | −2 |
| 101 | ILE | 0 | 258 | SAM | +1 |
| 118 | ASP | −1 | 267 | WAT | 0 |

**Table S2.** Structural details of the HcgC QM cluster model. The *holo* structure of HcgC used in this study is based on its parent PDB entries 5D4U and 5O4J with peer-reviewed publications[3,4]. The all-atom structure processed from the crystal structure was downloaded from their supporting information files of our previous QM/MM study on methyltransferases[2], from which the amino acid residues and small molecules in the QM region specified in the supporting information were extracted as the QM cluster model. Cleaved peptide bonds were capped with hydrogen atoms. The table lists the three-letter codes of the residues (Res. Name) included in the QM cluster model, along with their corresponding residue numbers in the reference full-protein structure (Res. Number) and net charges at their determined protonation states. In addition to standard three-letter codes, HIE means a neutral HIS where a proton resides at the $N_{\varepsilon}$ of the side chain, 9KH is the substrate (6-carboxy methyl-4-hydroxy-2-pyridinol), and SAM is the cofactor (*S*-adenosylmethionine). This leads to a QM cluster model with 545 atoms after capping and a net charge of −4.

| Res. Number | Res. Name | Net Charge | Res. Number | Res. Name | Net Charge |
|---|---|---|---|---|---|
| 1 | SAM | +1 | 208 | THR | 0 |
| 2 | 9KH | −3 | 209 | LEU | 0 |
| 7 | ILE | 0 | 238 | GLU | −1 |
| 8 | THR | 0 | 242 | PHE | 0 |
| 10 | SER | 0 | 300 | WAT | 0 |
| 58 | LYS | +1 | 301 | WAT | 0 |
| 80 | TYR | 0 | 304 | WAT | 0 |
| 82 | SER | 0 | 308 | WAT | 0 |
| 101 | ASP | −1 | 309 | WAT | 0 |
| 102 | ILE | 0 | 310 | WAT | 0 |
| 103 | GLN | 0 | 312 | WAT | 0 |
| 105 | HIE | 0 | 313 | WAT | 0 |
| 120 | LEU | 0 | 314 | WAT | 0 |
| 142 | THR | 0 | 317 | WAT | 0 |
| 143 | GLY | 0 | 318 | WAT | 0 |
| 144 | ILE | 0 | 327 | WAT | 0 |
| 145 | GLY | 0 | 329 | WAT | 0 |
| 146 | GLY | 0 | 330 | WAT | 0 |
| 147 | VAL | 0 | 331 | WAT | 0 |
| 148 | SER | 0 | 332 | WAT | 0 |
| 163 | GLU | −1 | 334 | WAT | 0 |
| 194 | THR | 0 | 335 | WAT | 0 |
| 206 | THR | 0 | 338 | WAT | 0 |
| 207 | MET | 0 | | | |

**Table S3.** QuantumPDB configurations for QM cluster construction of COMT enzymes. Key-value pairs in the config.yaml for QuantumPDB that affect the cluster construction result are listed in this table.

| Key | Value | Comment |
|---|---|---|
| modeller | true | Modeller[5] is used to preprocess the raw protein structure fetched from PDB. |
| protoss | true | Protoss[6] server is used to add hydrogens to the structure. |
| optimize_select_residue | 1 | Missing residues reported in the raw PDB file are built and optimized when Modeller preprocesses the raw structure. |
| max_clash_refinement_iter | 5 | The structure will be iteratively refined by Modeller and resent to the Protoss server when Protoss detects clashes between atoms after adding hydrogens until reaching 5 iterations. |
| max_atom_count | 300 | The maximum number of atoms is limited to 300 before capping the cleaved bonds. |
| number_of_spheres | 2 | The cluster includes at most the second coordination sphere of the central residues. |
| radius_of_first_sphere | 4.0 | The first coordination sphere is selected by geometric rules. Residues that have atoms within 4.0 Å of the central residues are included the first coordination sphere. |
| include_ligands | 2 | All atoms selected by the cluster construction algorithm are included in the cluster, whether they are amino acid residues, water molecules, ions, other small molecules, or other oligomers. |
| capping_method | 1 | The cleaved bonds in the cluster are capped with hydrogen. |
| smoothing_method | 2 | The outer coordination spheres are selected based on the Voronoi tessellation of the atomic structure. Dummy atoms are filled in the vacuum to regularize the shape of the Voronoi cells. |
| merge_distance_cutoff | 6.0 | Two central residues matched in the full structures are merged into the same center if their distance is within 6.0 Å. |

**Table S4.** Structural information for QM cluster construction of COMT enzymes. For these COMT models, we specified the Chemical Component Dictionary (CCD) code of a SAM cofactor, a ligand, and a $Mg^{2+}$ ion as the central residues in the QuantumPDB input to create the QM clusters. The CCD code of the ligand for each PDB ID is listed in the table. When multiple sets of central residues are matched from the full system structure, a cluster around one of them is picked for this study, as listed in the table by their chain IDs and residue IDs in the original PDB file. The number of atoms in the picked cluster after capping and the total charge of it, as counted by QuantumPDB, are also listed.

| PDB ID | Ligand | Picked Center Residues | #Atoms | Total Charge |
|---|---|---|---|---|
| 1VID | DNC | A300_A301_A302 | 318 | −1 |
| 2CL5 | BIE | B1216_B1217_B1218 | 318 | −1 |
| 2ZVJ | KOM | A300_A301_A302 | 303 | −1 |
| 3S68 | TCW | A222_A227_A228 | 303 | −1 |
| 4XUD | 43H | A301_A302_A303 | 320 | 0 |
| 4XUE | 43J | B301_B302_B303 | 310 | 0 |
| 5LSA | DNC | A301_A303_A304 | 310 | −1 |
| 6LFE | EAO | A301_A302_A303 | 301 | −1 |
| 7XJB | DNI | C301_C302_C303 | 313 | −1 |

**Table S5.** xTB and steered MD parameters for reactive sampling.

| Parameter | Value | Comment |
|---|---|---|
| Solvent | THF | ALPB implicit solvation in the xTB package. This solvent is chosen because it has a similar dielectric constant to the one used in the reference DFT calculations. |
| Thermostat | Langevin | ASE implementation |
| Time step | 0.5 fs | For the Langevin thermostat. |
| Friction coefficient | 0.01 $fs^{-1}$ | For the Langevin thermostat. |
| Temperature | 1000 K/500 K | For the Langevin thermostat. 1000 K is only for the PfPMT cluster. 500 K is for other clusters due to their larger size and more molecular fragments in the cluster. |
| Preparation time | 0.5 ps | |
| Production time | 12.5 ps | |
| Restraint $\kappa$ | 0.05 hartree/$Å^2$ | The spring constant of the Hookean restraint potential applied to all pairs between $C_\alpha$ of amino acid residues, water oxygens, and $Mg^{2+}$. |
| Bias potential $\kappa_0$ | 1000 kcal/mol/$Å^2$ | The spring constant of the MOVINGRESTRAINT harmonic potential in PLUMED. |
| Initial CV value | −2 Å | During the preparation run, the center of the bias potential moves from the CV value of the input structure to this value. |
| Final CV value | 2 Å | During the production run, the center of the bias potential moves from the CV from the initial value to the final value. |
| Sampling interval | 10 fs | Snapshots are only sampled during the production run. |

**Table S6.** Flexible scan settings in QM clusters. All scans are set between the CE atom of the unique SAM (residue ID: SAM) in the cluster and a nucleophilic atom (atom ID listed in the table) of the unique substrate (residue ID listed in the table). The target distances between the scanned atoms are also listed.

| System | Substrate Residue ID | Nucleophilic Atom ID | Target Distance (Å) |
|---|---|---|---|
| PfPMT | OPE | N | 1.492 |
| HcgC | 9KH | C5 | 1.500 |
| COMT-2ZVJ | KOM | O9 | 1.430 |
| COMT-4XUD | 43H | O | 1.430 |
| COMT-4XUE | 43J | O | 1.430 |
| COMT-6LFE | EAO | O19 | 1.430 |
| COMT-7XJB | DNI | O24 | 1.430 |
| COMT-1VID | DNC | O2 | 1.430 |
| COMT-2CL5 | BIE | O2 | 1.430 |
| COMT-3S68 | TCW | O8 | 1.430 |
| COMT-5LSA | DNC | O2 | 1.430 |

**Table S7.** Architectural configurations of Enerzyme-PhysNet. Our Enerzyme-PhysNet core is a reimplementation in PyTorch of the original TensorFlow version in the official PhysNet GitHub repository[11]. We used Distance, Range Separation, Exponential Gaussian RBF, and Random atom embedding pre-core layers, and Readout, Atomic Affine, Charge conservation, Electrostatic energy (PhysNet flavor), Atomic charge to dipole, Energy reduce, and Force post-core layers (Table S11). We adopted the recommended model depth and width of the PhysNet and SpookyNet in their original literature.

| **Key** | **Value** |
|---|---|
| Short-range cutoff radius | 10 Å |
| Long-range cutoff radius | $+\infty$ |
| Dimension of the atom type embedding | 128 |
| Number of radial basis functions | 64 |
| Number of interaction blocks | 5 |
| Number of residual layers for the atomic embedding preprocessing | 2 |
| Number of residual layers in the interaction block | 3 |
| Number of residual layers in the output block | 1 |

**Table S8.** Architectural configurations of Enerzyme-SpookyNet. Our Enerzyme-SpookyNet core customized the neural network layer initialization of the official SpookyNet GitHub repository[12] for better training dynamics. We used Distance, Range Separation, Exponential Bernstein RBF, Nuclear embedding, and Electronic embedding pre-core layers, and Readout, Atomic Affine, Charge conservation, Electrostatic energy (SpookyNet flavor), Atomic charge to dipole, Energy reduce, and Force post-core layers (Table S11). We adopted the recommended model depth and width of the PhysNet and SpookyNet in their original literature.

| **Key** | **Value** |
|---|---|
| Short-range cutoff radius | 5 Bohr |
| Long-range cutoff radius | $+\infty$ |
| Dimension of the atom type embedding | 128 |
| Number of radial basis functions | 16 |
| Number of interaction blocks | 6 |
| Number of residual layers for the atomic embedding preprocessing | 1 |
| Number of residual layers in the interaction block | 1 |
| Number of residual layers for the atomic feature postprocessing | 1 |
| Number of residual layers in the output block | 1 |

**Table S9.** Architectural configurations of Enerzyme-MACE. Our Enerzyme-MACE core borrows the official MACE GitHub repository[13] and chooses the MACE-OFF-M architecture[14]. We used Distance, Range Separation, Exponential Bernstein RBF, Nuclear embedding, and Electronic embedding pre-core layers, and Readout, Atomic Affine, Charge conservation, Electrostatic energy (PhysNet flavor), Atomic charge to dipole, Energy reduce, and Force post-core layers (Table S11).

| Key | Value |
|---|---|
| Short-range cutoff radius | 5 Å |
| Long-range cutoff radius | $+\infty$ |
| Dimension of the atom type embedding | 128 |
| Number of radial basis functions | 8 |
| Number of interaction blocks | 2 |
| Maximum angular momentum | 3 |
| Maximum order of correlation | 3 |
| Dimension of additional hidden irreps | 128×1o |
| Dimension of radial MLP | [64, 64, 64] |
| Dimension of MLP irreps | 16×0e |
| Gate activation function | silu |
| Averaged number of neighbors | 23.33 |

**Table S10.** Default Training hyperparameters of Enerzyme-NNPs.

| Hyperparameter Description | Hyperparameter Value | | |
|---|---|---|---|
| | PhysNet | SpookyNet | MACE |
| Learning rate | 0.001 | 0.001 | 0.01/0.001 |
| Weight decay | 0 | 0 | 0 |
| Adam optimizer's beta | (0.9, 0.999) | (0.9, 0.999) | (0.9, 0.999) |
| Adam optimizer's eps | $10^{-6}$ | $10^{-6}$ | $10^{-6}$ |
| Batch size | 8 | 8 | 10 |
| Maximum epochs | 10000 | 10000 | 10000 |
| Early stopping patience | 200 | 200 | 200 |
| Warm-up epochs | 10 | 10 | 10 |
| EMA decay rate | 0.999 | 0.999 | 0.999 |
| Floating point data type | FP32 | FP32 | FP32 |
| Energy weight $w_E$ | 1 | 1 | 1 |
| Forces weight $w_{\boldsymbol{F}}$ | 1000 | $1000\sqrt{3}$ | 1000 |
| Dipole moment weight $w_\mu$ | 1 | 1 | 1 |
| Total charge weight $w_Q$ | 1 | 1 | 1 |
| Atomic charges weight $w_q$ | 100 | 100 | 100 |
| Loss type $\ell_x$ | MAE | RMSE | MAE |
| Metric type $m_x$ | RMSE | RMSE | RMSE |

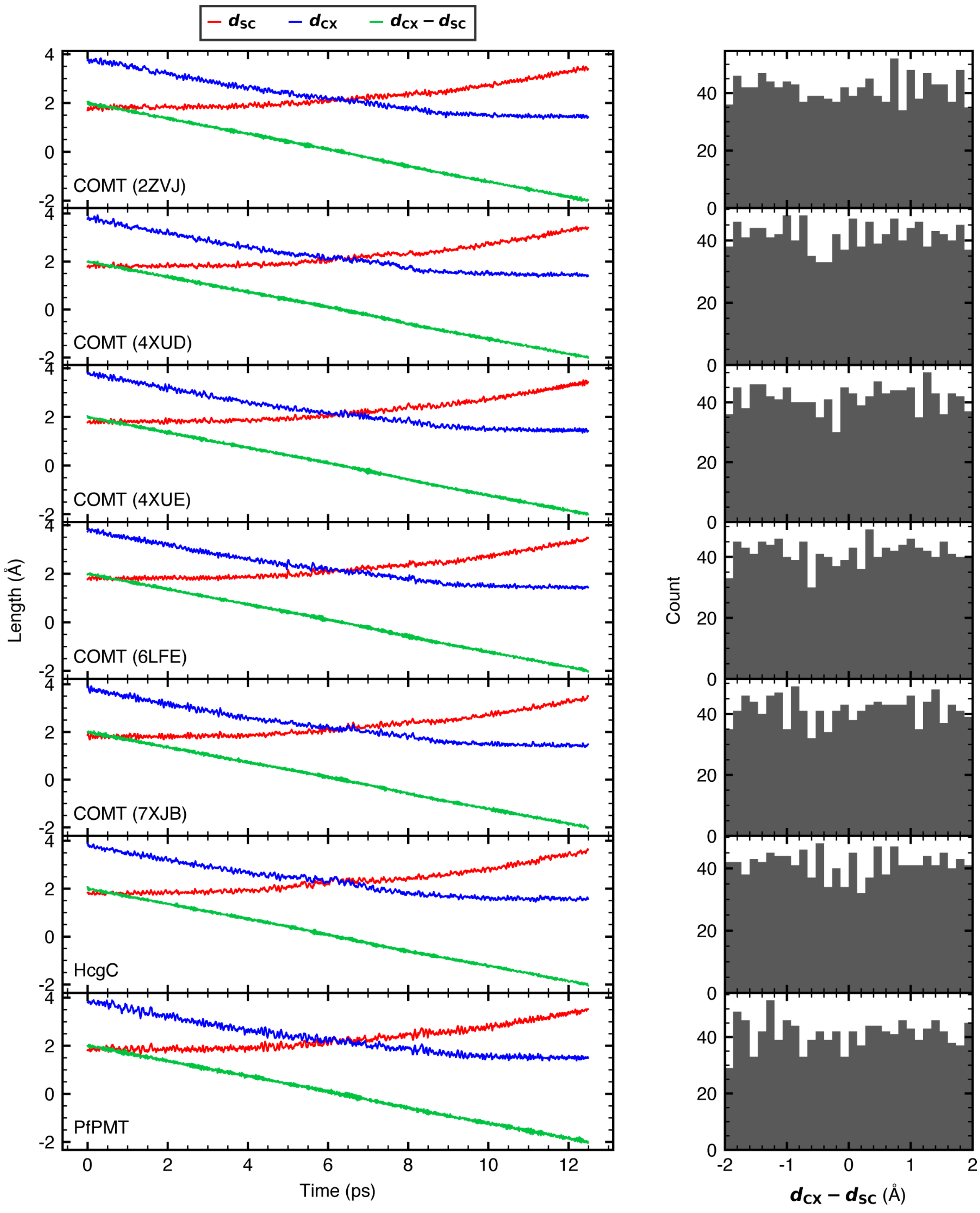

**Figure S1.** The sampling outcome of reaction coordinates over steered MD trajectories. The left column is the time evolution of reaction coordinates throughout the 12.5 ps steered MD production run. The distances $d_{CX}$ and $d_{SC}$ are defined in Figure 3 in the main text. The right column contains histograms of the $d_{CX}$–$d_{SC}$ value over the 1,250 snapshots used as NNP datasets. For the 5 COMT systems, their structure sources are annotated with their PDB IDs.

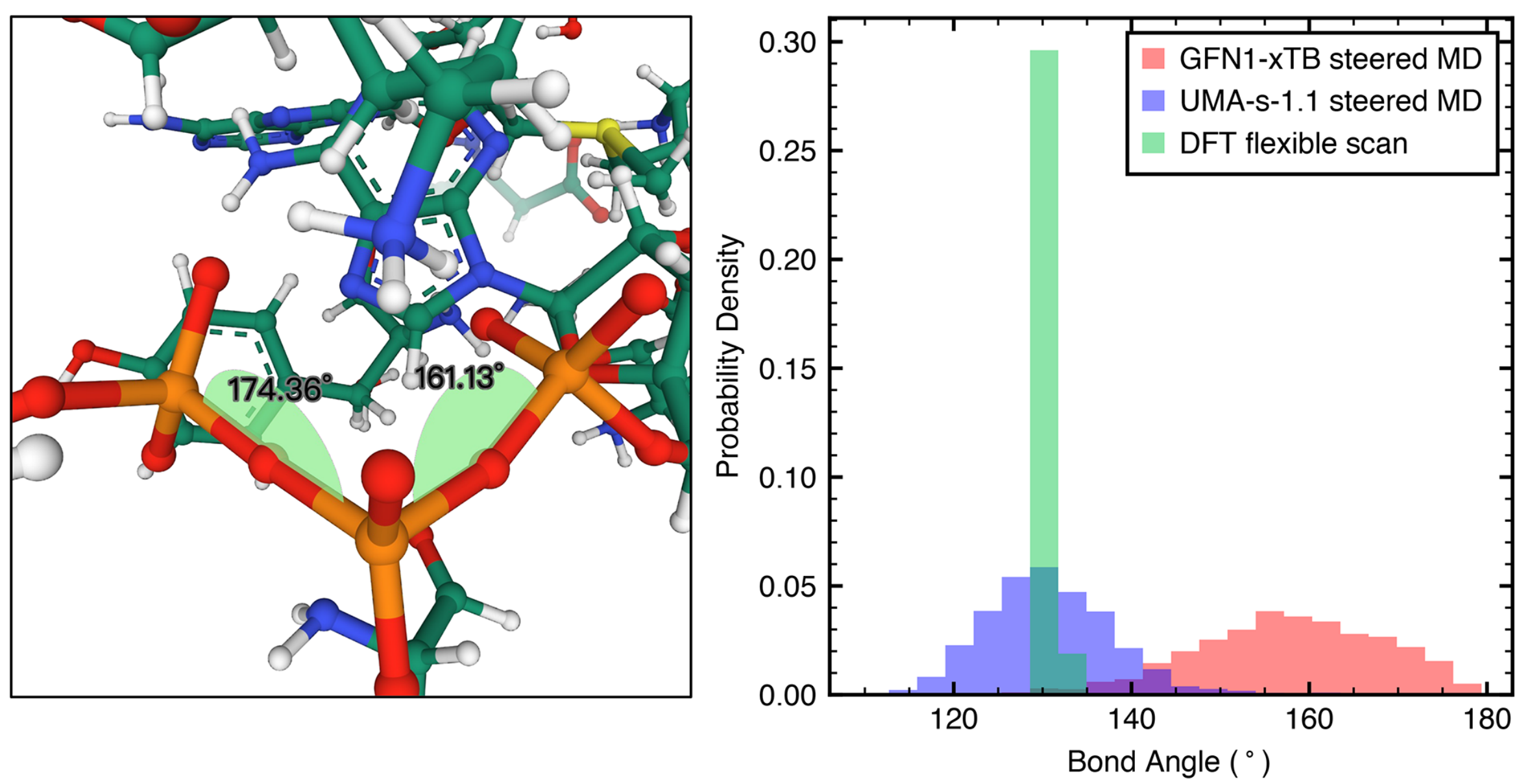

**Figure S2.** Triphosphate structures at different levels of theory. Left: This snapshot shows a common pattern of unphysical triphosphate structures encountered in steered MD driven at the GFN1-xTB level of theory. In this 301-atom cluster extracted from the SARS-CoV-2 nsp16-nsp10 2'-*O*-methyltransferase structure (PDB ID: 6WVN) around its substrate analog 7-methyl-GpppA (CCD code: GTA) and SAM (A7112_A7113) by the standard QuantumPDB procedure used for COMTs in this study, starting from atomic positions in the crystal structure, the triphosphate group tends to manifest unphysically large P–O–P bond angles during the standard steered MD production run at 500 K used for all other clusters in this study. Right: The histograms of the triphosphate P–O–P bond angles over the standard steered MD production runs of the 6WVN cluster driven by GFN1-xTB semi-empirical quantum chemistry and UMA-s-1.1 universal NNP, respectively. A histogram from a 25-step flexible scan between the carbon of the transferring methyl group and the 2’ oxygen of the GTA (O2B) from the optimized distance of the crystal structure to 1.48 Å, at the standard DFT level of theory used for all other clusters in this study, is also included as a reference.

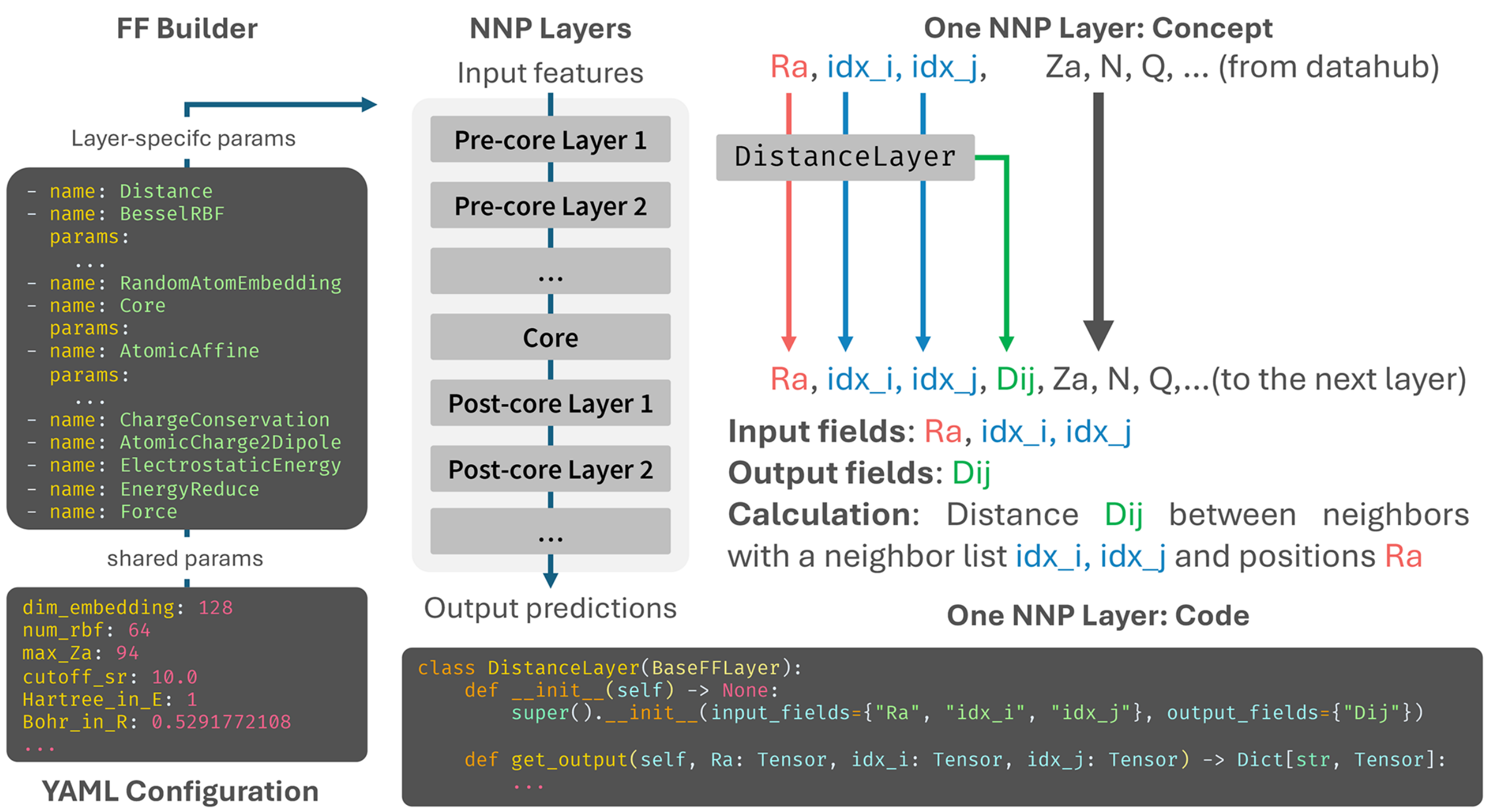

**Figure S3.** The modular design of NNPs in Enerzyme. In Enerzyme, every NNP is disassembled into sequential neural network layers. The FF Builder in Enerzyme initializes those layers sequentially and loads each layer's shared and specific parameters from a YAML configuration file to build a fully assembled NNP. Information flow across layers is organized as a dictionary of fields that store different features, predictions, or intermediate variables of the molecule. Each layer of the NNP runs the calculation for its own output fields only with its designated input fields from the previous layer's outputs and directly passes other fields forward. These data-passing logistics are encapsulated within the BaseFFLayer abstract class, and its subclass layers only implement the algorithmic part.

**Table S11.** Pre-core and Post-core layers in Enerzyme.

| Layer | Type | Functionality |
|---|---|---|
| Distance | Pre-core | Compute distances and translation vectors for all edges in the molecular graph. |
| Range separation | Pre-core | Separate short-range edges with a distance larger than a given cutoff, and those with a distance no larger than the cutoff, into different fields. |
| Exponential Gaussian RBF | Pre-core | Compute the exponential Gaussian radial basis function (RBF) used in the original PhysNet[7] from the distances for all short-range edges in the molecular graph. |
| Exponential Bernstein RBF | Pre-core | Compute the exponential Bernstein polynomial RBF used in the original SpookyNet[8] from the distances for all short-range edges in the molecular graph. |
| Bessel RBF | Pre-core | Compute the exponential Bessel function RBF used in the original MACE[9] from the distances for all short-range edges in the molecular graph. |
| Random atom embedding | Pre-core | Embed atom types into learnable random vectors used in the original PhysNet[7]. |
| Nuclear embedding | Pre-core | Embed atom types into learnable random vectors fused with a learnable transformation on ground-state electronic configuration features of the atom type, introduced in SpookyNet[8]. |
| Electronic embedding | Pre-core | Embed the total charge or the total spin multiplicity of the system into learnable vectors for each atom, introduced in SpookyNet[8]. |
| Gather atom embedding | Pre-core | Gather all atom embedding vectors with a summation that obtains the embeddings ready to be passed into the message passing core. |
| Readout | Post-core | Read out atomic properties like energies or charges by applying NN block to the atom features from the message passing core |
| Atomic affine | Post-core | Do element-wise, learnable shift and scale on atomic properties, used in the original PhysNet[7] and SpookyNet[8]. |
| Charge conservation | Post-core | Correct the predicted atomic charges with the input total charge by a linear transformation, used in the original PhysNet[7] and SpookyNet[8]. |
| Electrostatic energy | Post-core | Compute the Coulomb potential based on the system geometry and predicted atomic charges, with different damping flavors in the original PhysNet[7] and SpookyNet[8]. |
| Atomic charge to dipole | Post-core | Compute the system dipole moment based on the system geometry and predicted atomic charges. |
| Energy reduce | Post-core | Reduce the atomic energies to the total energy of the system |
| Force | Post-core | Compute the forces on each atom through auto-differentiation of the total energy with respect to the coordinates. |

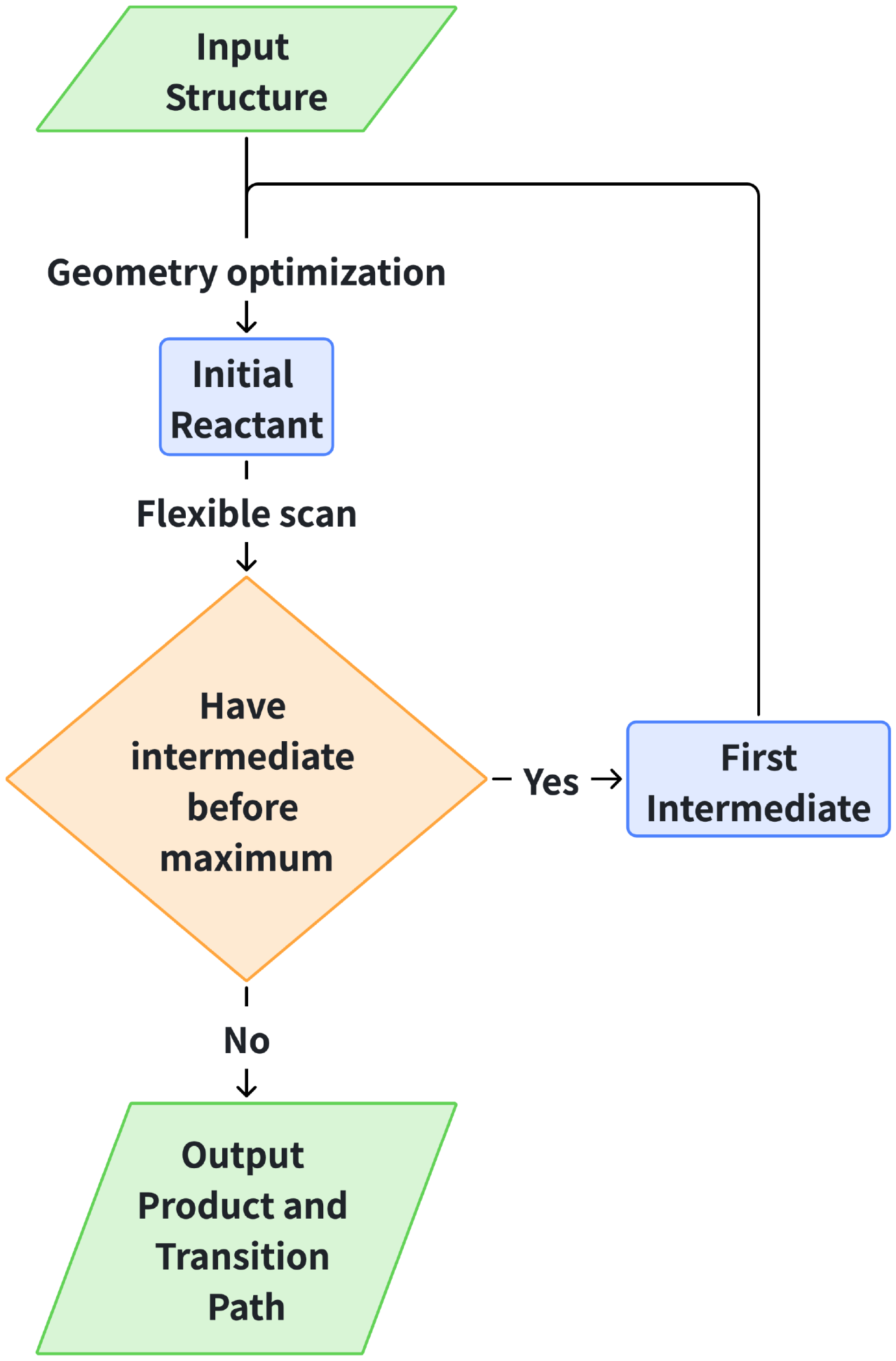

**Figure S4.** Automated flexible scan workflow in Enerzymette. This flowchart shows the iterative NEB algorithm in the Enerzymette `enerzyme_scan` submodule. The highest-energy structure of the output transition path is treated as the estimated transition state (TS). All optimized reactants and products, including the original endpoints and the newly discovered intermediates, are also collected.

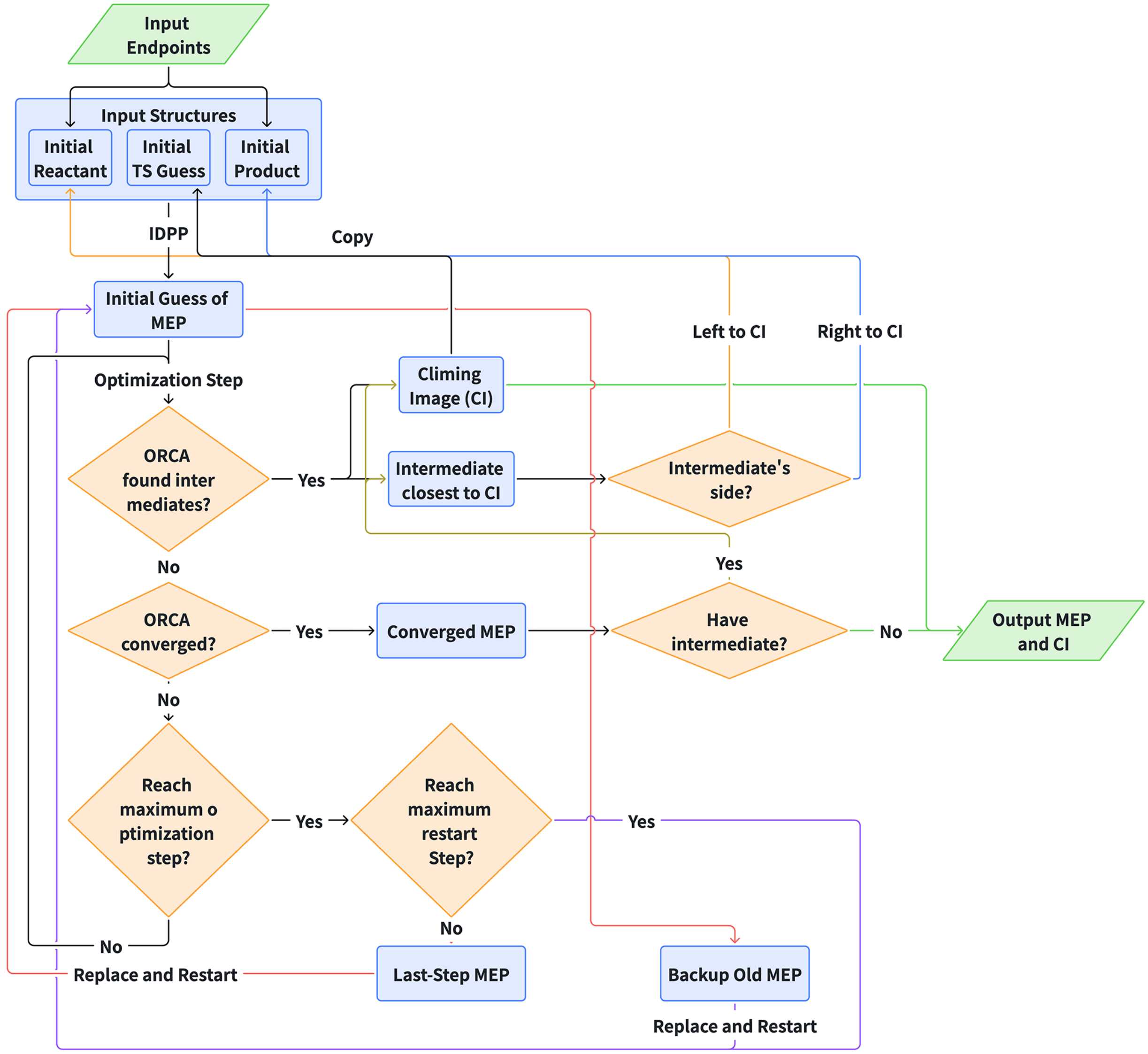

**Figure S5.** Automated CI-NEB workflow in Enerzymette. This flowchart shows the iterative NEB algorithm in the Enerzymette `enerzyme_neb` submodule. The output climbing image (CI) is treated as the estimated transition state (TS). All optimized reactants and products, including the original endpoints and the newly discovered intermediates, are also collected. IDPP stands for image dependent pair potential, an interpolation scheme, and MEP stands for minimum energy pathway.

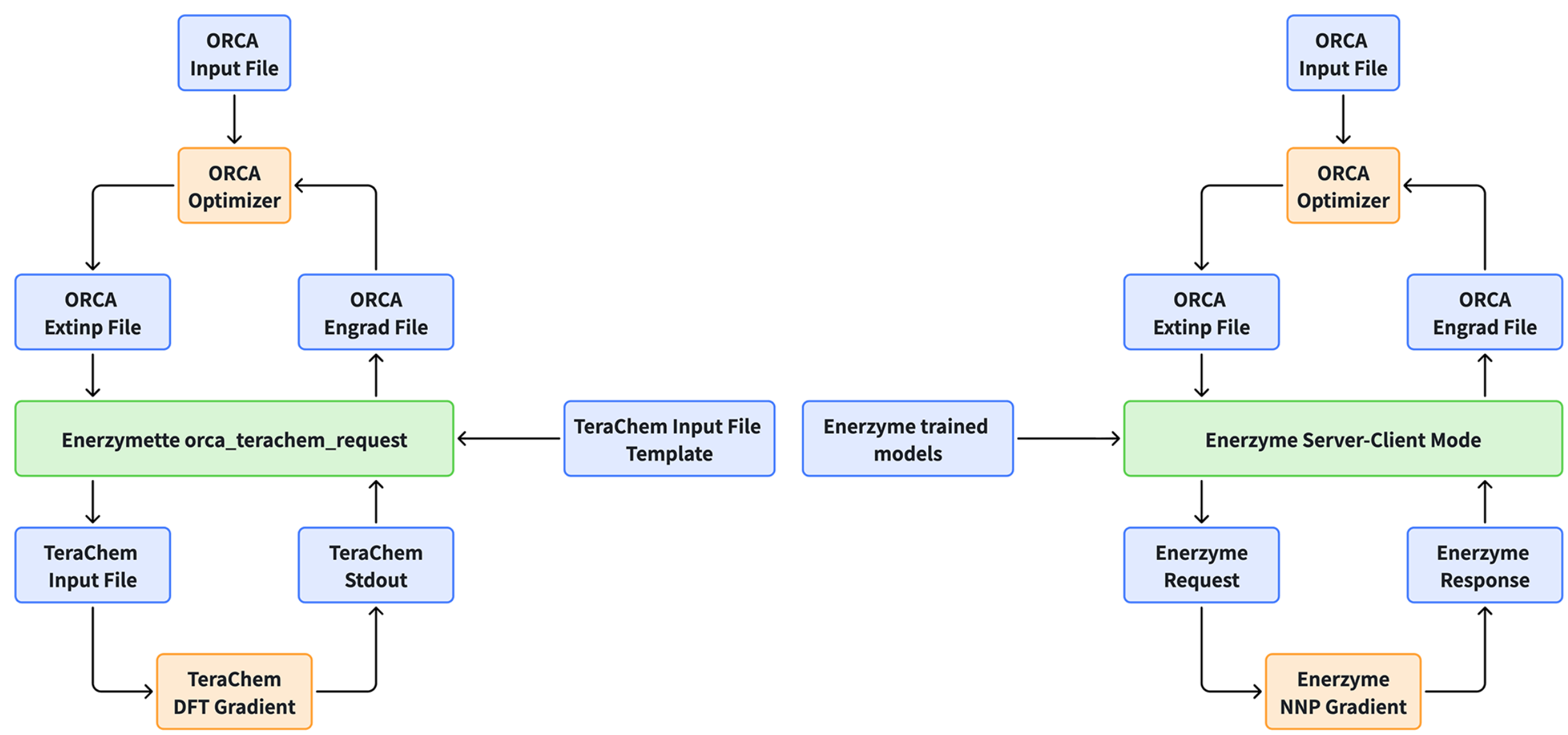

**Figure S6.** The ORCA-TeraChem and ORCA-Enerzyme interfaces. These flowcharts show how ORCA, as an external optimizer, interfaces with TeraChem via the Enerzymette orca_terachem_request submodule or with Enerzyme via its server-client mode.

**Table S12.** NEB-Estimated reaction energetics in HcgC and PfPMT cluster models. $\Delta E$ is the estimated reaction energy change, and $\Delta E^{\ddagger}$ is the estimated reaction kinetic barrier.

| **System** | **Method** | **$\Delta E$ (kcal/mol)** | **$\Delta E^{\ddagger}$ (kcal/mol)** |
|---|---|---|---|
| HcgC | Reference DFT | -18.1 | 20.7 |
| | Enerzyme-SpookyNet | -17.7 | 18.3 |
| | Enerzyme-MACE | -11.8 | 17.1 |
| PfPMT | Reference DFT | -9.7 | 23.1 |
| | Enerzyme-SpookyNet | -7.8 | 22.8 |
| | Enerzyme-MACE | -11.1 | 26.2 |

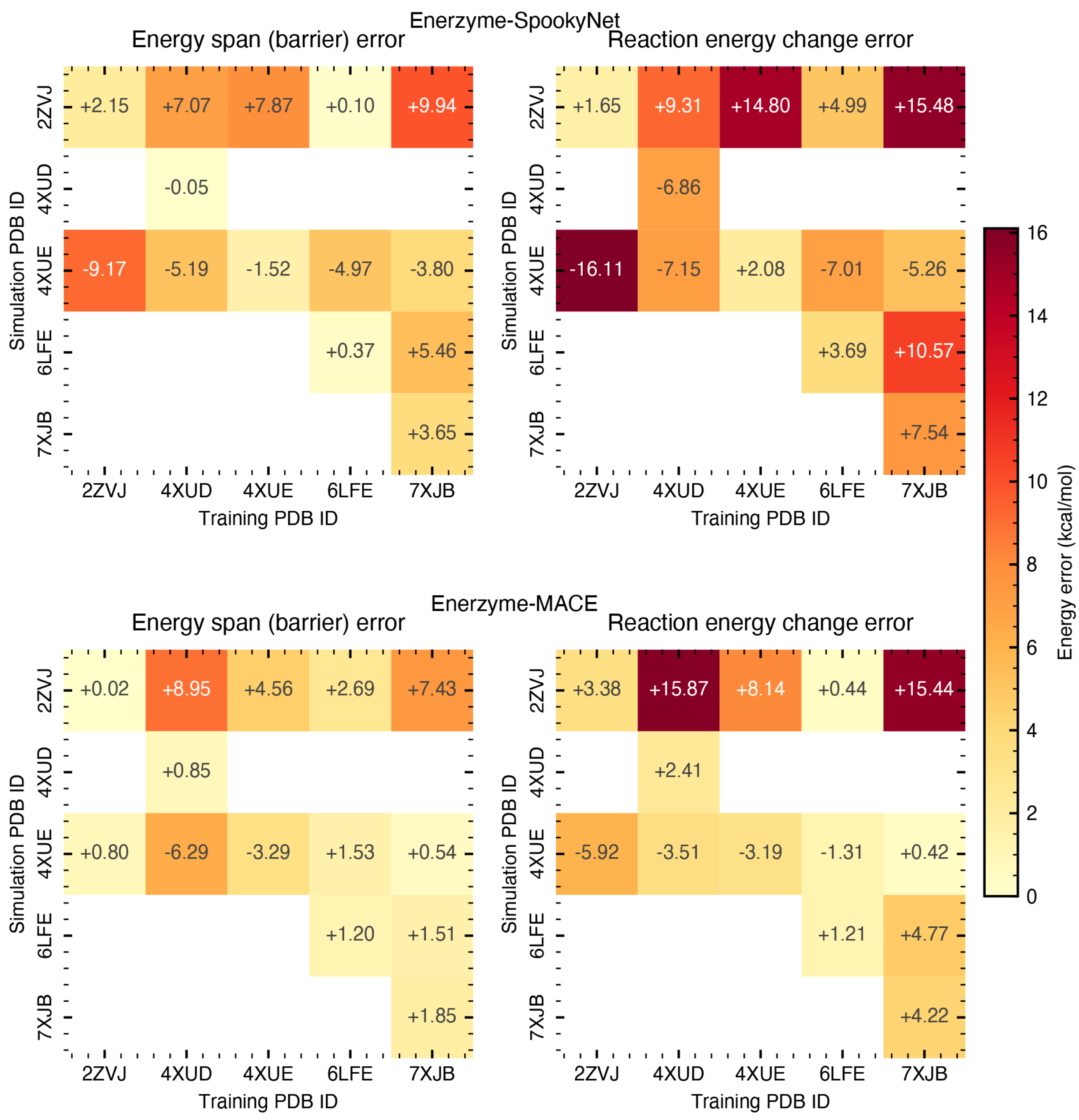

**Figure S7.** The error of the energy span (kinetic barrier) and reaction energy change error from iterative scan simulations when the Enerzyme-NNP models are trained on a COMT cluster of the training PDB ID and evaluated on a COMT cluster of the simulation PDB ID.

**Table S13.** Global geometry accuracy of Enerzyme-NNPs on reactants, products and TS from NEB in cluster models. The RMSD is the root mean square deviation of all unconstrained atoms between NNP- and DFT- optimized substructures.

| System | Method | Reactant RMSD (Å) | Product RMSD (Å) | TS RMSD (Å) |
|---|---|---|---|---|
| HcgC | Enerzyme-SpookyNet | 0.21 | 0.18 | 0.19 |
| | Enerzyme-MACE | 0.41 | 0.11 | 0.40 |
| PfPMT | Enerzyme-SpookyNet | 0.46 | 0.58 | 0.48 |
| | Enerzyme-MACE | 0.46 | 0.28 | 0.40 |

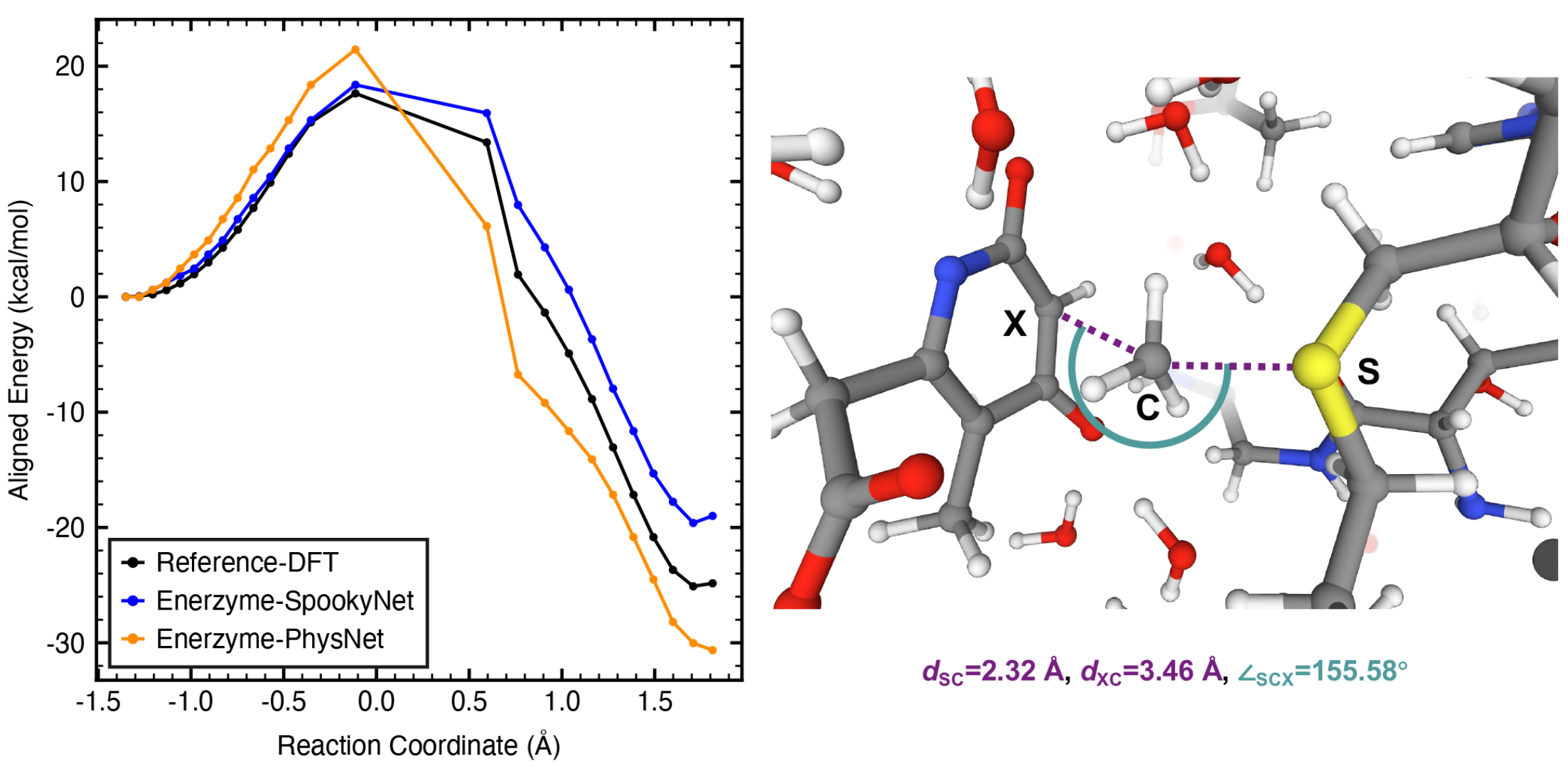

**Figure S8.** The failure pattern of Enerzyme-PhysNet in the HcgC cluster NEB. Left: Single-point energy evaluation of Enerzyme-SpookyNet and Enerzyme-PhysNet along DFT flexible scan trajectories in the HcgC cluster model. The initial energies of different energy curves are aligned to be the same. Right: the distorted TS structure estimated by the iterative NEB with Enerzyme-PhysNet with a too-small bond angle and a too-large $d_{CX}$ as defined in Figure 3 in the main text. Color scheme: carbon in gray, hydrogen in white, oxygen in red, nitrogen in blue, and sulfur in yellow.

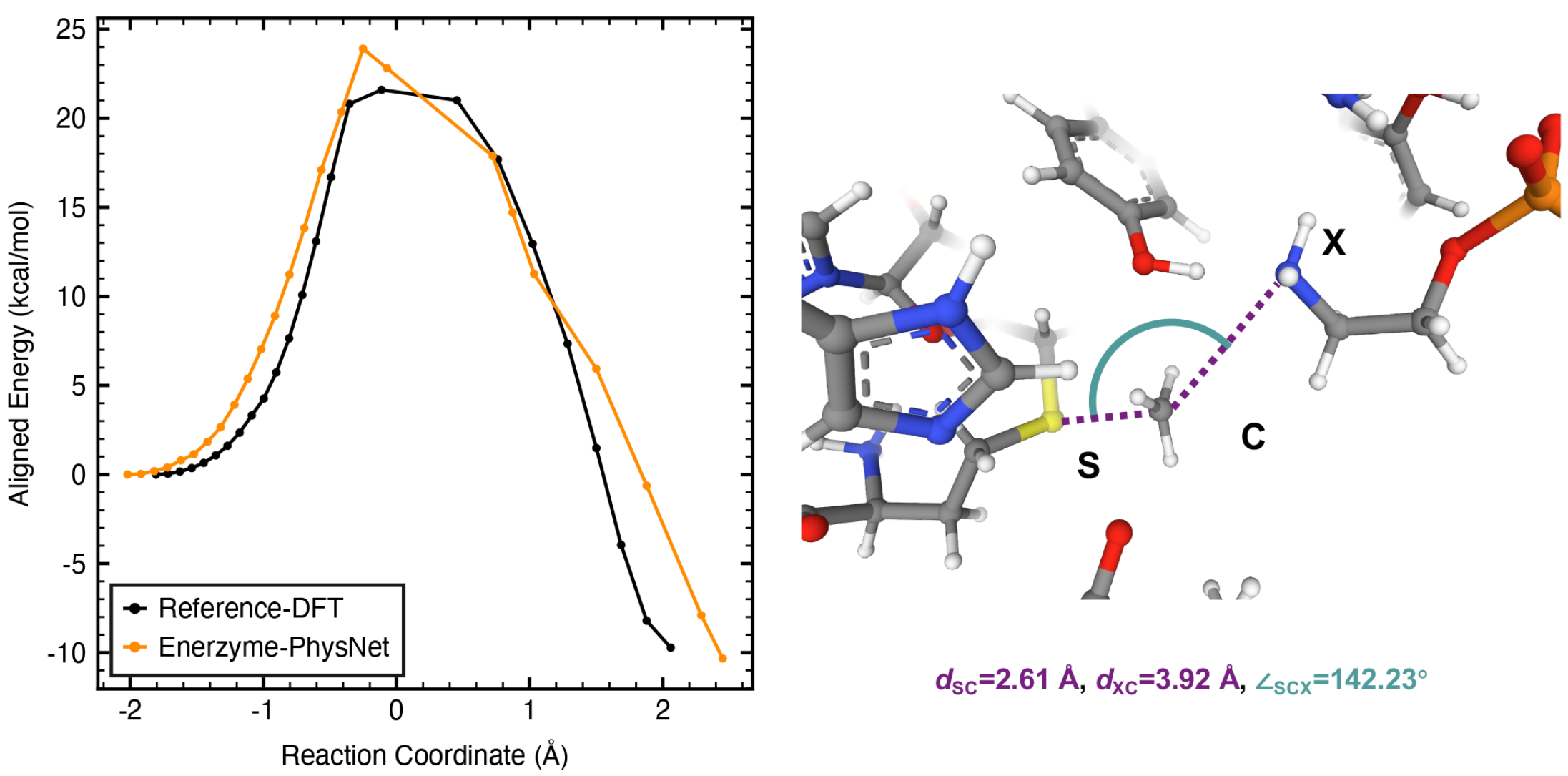

**Figure S9.** The failure pattern of Enerzyme-PhysNet in the PfPMT cluster NEB. Left: Energy profile of the flexible scan of the PfPMT cluster model with Enerzyme-PhysNet. The initial energies of different energy curves are aligned to be the same. Right: the distorted TS structure estimated by the iterative NEB with Enerzyme-PhysNet with a too-small bond angle and a too-large $d_{CX}$ as defined in Figure 3 in the main text. Color scheme: carbon in gray, hydrogen in white, oxygen in red, nitrogen in blue, and sulfur in yellow.

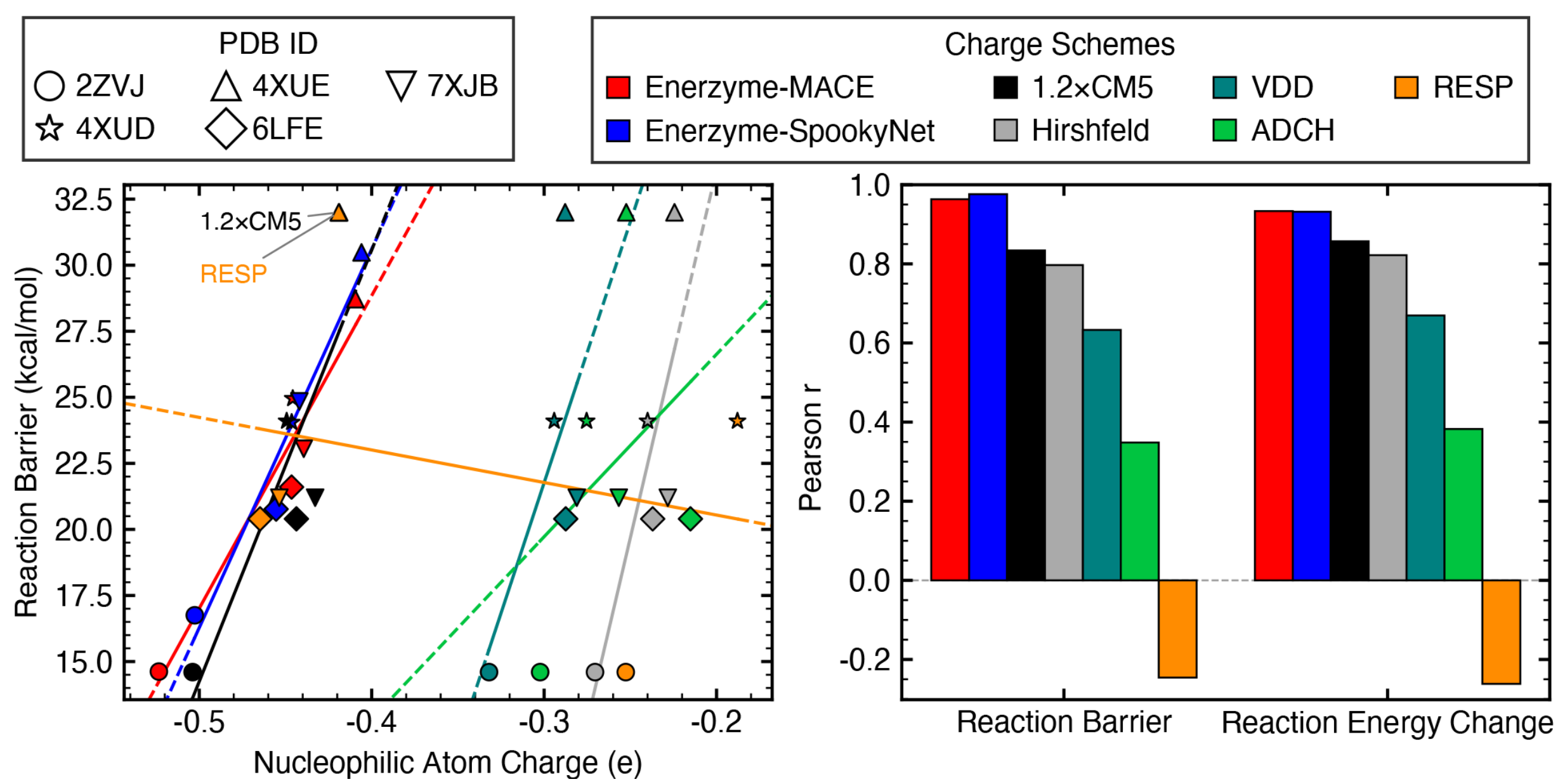

**Figure S10.** Charge-reactivity correlation for more atomic charge schemes. The correlation between the partial charge of the nucleophilic atom of COMT substrates and the reaction barrier and the reaction energy change. For all systems from 5 PDB IDs (Scheme 1 in the main text), their reaction barriers and atomic charges obtained from reference DFT flexible scan with electronic structure analysis and Enerzyme NNP flexible scan with atomic charge prediction are plotted at the top left. Left: The lines show the linear regression between the predicted reaction barrier of the 5 systems and the predicted nucleophilic atom charge with their corresponding markers for different PDB IDs. 1.2xCM5 and RESP overlap for one point, as indicated by inset labels, and the remaining charge schemes are colored according to the legend shown at top. Right: The Pearson *R* coefficients for the linear regression between the predicted reaction barrier/reaction energy change for the 5 systems and the predicted nucleophilic atom charge across different charge schemes.

| **Ligand** | | | |
|---|---|---|---|
| **PDB ID** | **1VID, 5LSA** | **2CL5** | **3S68** |

**Figure S11.** Ligand structures and their associated PDB IDs for the protein-ligand complex crystal structures of the 4 out-of-dataset COMT systems used in the Enerzyme-NNP cross-system evaluations. The nucleophilic atoms in the methyl transfer reactions are colored.

**Table S14.** Mean absolute error (MAE) of reaction barriers obtained by flexible scan with Enerzyme-NNPs trained on COMT datasets compared to reference DFT results. Data ratio 1× stands for 175 datapoints per system for 5 COMT systems in the dataset. In-dataset metrics are evaluated on the 5 COMT systems in the dataset (Scheme 1 in the main text), while out-of-dataset metrics are evaluated on other 4 COMT systems in the dataset (Figure S11).

| Data ratio | Method | In-dataset MAE (kcal/mol) | Out-of-dataset MAE (kcal/mol) |
|---|---|---|---|
| 1× | Enerzyme-SpookyNet | 3.24 | 3.22 |
| | Enerzyme-MACE | 2.18 | 1.99 |
| 2× | Enerzyme-SpookyNet | 2.08 | 2.31 |
| | Enerzyme-MACE | 1.79 | 1.79 |
| 4× | Enerzyme-SpookyNet | 1.74 | 1.72 |
| | Enerzyme-MACE | 1.32 | 1.98 |